\documentclass[pdflatex,sn-mathphys-ay]{sn-jnl}

\usepackage{amssymb}
\usepackage{amsmath}
\usepackage{enumitem}
\def\T{{\mathrm{\scriptscriptstyle T}}}
\newcommand{\suppref}[1]{\hyperref[#1]{\ref*{#1}}}
\newcounter{suppsec}
\renewcommand{\thesuppsec}{section S\arabic{suppsec}}
\newcommand{\suppsection}[2]{%
    \refstepcounter{suppsec}%
    \hypertarget{#1}{}\textbf{\thesuppsec:} #2\label{#1}%
}
\setcitestyle{aysep={,}}
\usepackage{pdfpages}

\begin{document}
	
\title{Inference and local influence diagnostics for unit-Lindley additive partially linear models}
	
	
\author[1]{\fnm{Hatice Tul Kubra} \sur{Akdur}}\email{haticesenol@gazi.edu.tr}

\author[2]{\fnm{Danilo V.} \sur{Silva}}
\email{danilo.silva@ime.usp.br}

\author*[2]{\fnm{Gilberto A.} \sur{Paula}}
\email{giapaula@ime.usp.br}

\affil[1]{\orgdiv{Department of Statistics, Faculty of Science}, \orgname{Gazi University}, \orgaddress{\city{Ankara}, \country{Turkey}}}
	
\affil[2]{\orgdiv{Department of Statistics, Institute of Mathematics, Statistics and Computer Science}, \orgname{Universidade de S\~ao Paulo}, \orgaddress{\city{S\~ao Paulo}, \country{Brazil}}}

	
\abstract{This paper introduces a novel regression framework for modeling response variables restricted to the unit interval by proposing unit-Lindley additive partially linear models (UL-APLMs). This model class combines parsimony and interpretability of one-parameter unit-Lindley distribution with the flexibility of additive partial linear structures, enabling the coexistence of linear and smooth covariate effects. Additive terms are modeled using B-spline basis under a penalized likelihood framework to ensure smoothness. Estimation is carried out by maximizing the penalized log-likelihood function. The goodness-of-fit of the models is assessed through residual analysis, whereas the robustness of the parameter estimates and the detection of influential data points are evaluated using the local influence approach, which incorporates curvature diagnostics under case-weight and response perturbation schemes. The sensitivity and penalized observed information matrices are derived explicitly for the proposed model. Simulation studies demonstrate the accuracy of the estimation procedure under various scenarios. Real data on the assessment of the psychological profile of patients with hypopituitarism illustrate the applicability of the model, highlighting the diagnostic importance.}

	
\maketitle
\section{Introduction} 
\label{sec_intro}

Modeling response variables bounded in the unit interval, such as rates and proportions, has long attracted attention in the statistical literature due to their prevalence in many fields, including medicine, ecology, and economics. Traditional models for such data often rely on the beta distribution, which offers flexibility but can be computationally intensive and, in some cases, lead to parameter estimation issues. As an alternative, \citet{mazucheli_2019} introduced the one-parameter unit-Lindley (UL) distribution, a simpler, tractable distribution defined on the unit interval that captures skewed behavior with fewer parameters. The unit-Lindley distribution, derived from a transformation of the Lindley distribution, exhibits an appealing exponential form and closed-form expressions for its cumulative distribution function and moments, making it well-suited for modeling bounded outcomes with moderate skewness. The UL regression model for rates and proportions was developed for independent responses, and \citet{silva_2023} recently extended it to the GEE structure for analyzing cluster data and longitudinal and repeated-measure studies under the marginal unit-Lindley distribution.

In recent years, there has been an increasing interest in combining linear and additive terms within a unified regression framework to flexibly capture complex relationships in data. Partial linear models and their additive extensions are one such class, allowing for linear effects of some covariates and nonlinear (often smooth) effects of others. These models are especially useful when it is difficult to specify the functional form of some covariates initially. The penalized likelihood framework has been widely employed for estimation in these settings, often using B-spline bases and roughness penalties \citep{eilers_1996}. This framework enables control over the trade-off between model fit and smoothness.

While several distributions have been incorporated into additive models, including the Gaussian, gamma, and beta distributions, there has been little exploration of such models under the unit-Lindley distribution. This paper proposes an UL-APLM that extends the work of Mazucheli et al. (2019) by allowing both linear and smooth effects of explanatory variables. Estimation is performed via maximization of a penalized log-likelihood, where additive terms are modeled with B-splines and penalized to ensure smoothness.

Residual analysis for UL-APLMs is performed by deriving the quantile residual \citep{dunn_1996} for the UL distribution, and to assess the robustness of the parameter estimates and identify influential observations, we further conduct a local influence analysis based on the methodology introduced by \citet{cook_1986}, extended to additive frameworks by various authors \citep{hart_lof, xie_2007}. This technique allows the examination of small perturbations in data or model structure and their impact on estimation. Specifically, we investigate case-weight and response perturbation schemes and adapt curvature-based influence diagnostics to the unit-Lindley framework. The derivations build on sensitivity matrix theory and penalized observed information matrix, following approaches similar to those employed in GAM \citep{green_glm}. 

Our study contributes to the statistical modeling literature by incorporating a novel, computationally efficient distribution into the flexible additive modeling paradigm, thereby addressing both distributional and functional complexities in bounded-response data. Through simulation studies and a real-data application, we demonstrate the model's interpretability, practical performance, and diagnostic capabilities.

The paper is organized as follows. In \autoref{sec_iter_proc}, the UL-APLMs are introduced with some inferential results and the iterative process to obtain the MPLEs. The \autoref{sec_diag} extends several diagnostic procedures to the UL-APLM framework, including residual analysis and sensitivity assessment based on the local influence approach, as well as the computation of effective degrees of freedom and the estimation of the smoothing parameter. Simulation studies are conducted in \autoref{sec_simu} to assess the distribution of the estimators' samples. A real dataset is analyzed in \autoref{sec_aplic} and the final section presents some concluding remarks. The R functions and dataset used in the paper are available as supplementary materials.

\section{Penalized B-spline} \label{sec_iter_proc}
		
We adapt the methodology of additive partially linear models with B-spline smoothing as described in the context of the generalized log-gamma model \citep{cardozo_2022} to the unit-Lindley regression framework. We consider an additive partially linear structure in which both linear and smooth additive components are present.
				
\subsection{The model}
\begin{figure}[htb]
\centering
\includegraphics[width=13cm]{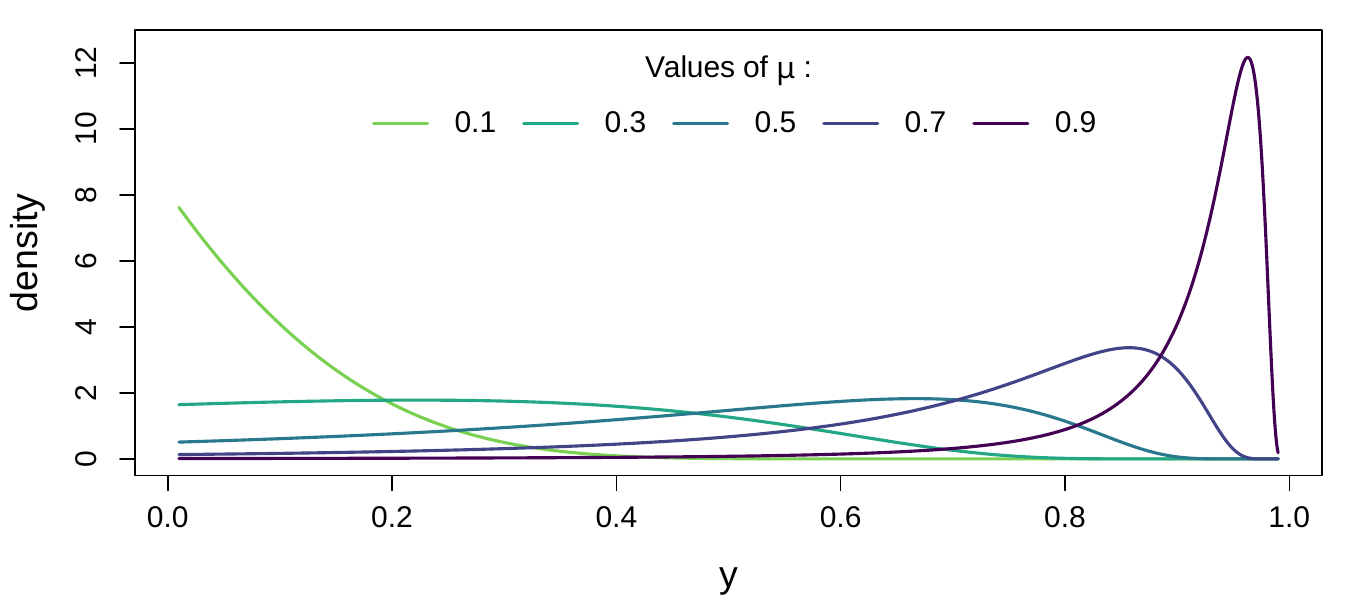}
\caption{Probability density function of the unit-Lindley distribution for different mean values.}
\label{bsplie_densi}
\end{figure}

Let $Y_i \stackrel{\rm ind} {\sim} \text{UL}(\mu_i)$ for $i=1,\ldots,n$, the unit-Lindley density is
\[
f(y_i \mid \mu_i) = \frac{(1 - \mu_i)^2}{\mu_i (1 - y_i)^3} 
\exp\left\{-\frac{y_i (1 - \mu_i)}{\mu_i (1 - y_i)}\right\}, 
\quad 0 < y_i < 1, \quad 0 < \mu_i < 1.
\]
The probability density in \autoref{bsplie_densi} illustrates the $\text{UL}(\mu)$ distribution for several values of the mean parameter $\mu$. We model the mean parameter $\mu_i$ via the relation
\begin{equation} \label{eq_predictor}
g(\mu_i) = {\bf x}_i^\T {\boldsymbol \beta} + \sum_{j=1}^r f_j(t_{ij}),
\end{equation}
where $g(\cdot)$ is a strictly monotonic and twice differentiable link function mapping $(0,1)$ to $\mathbb{R}$, ${\bf x}_i \in \mathbb{R}^p$ is a vector of covariate values with associated linear term coefficient vector ${\boldsymbol \beta} = (\beta_1, \ldots, \beta_p)^\T$, and $f_j(\cdot)$ are unknown smooth functions of continuous covariates $t_{ij}$. Each smooth term $f_j(\cdot)$ is represented using a \(q_j\)-dimensional B-spline basis. For independent observations, \( Y_i \ {\sim} \ \text{UL}(\mu_i) \) with the mean parameter modeled using a predictor as in \eqref{eq_predictor}. Under this specification, the model assumes that the conditional mean of the response variable satisfies
\begin{equation} \label{eq_mean}
\mu_i = {\mathbb E}(Y_i) = g^{-1} \left\{ {\bf x}_i^\T {\boldsymbol \beta} + \sum_{j=1}^r f_j(t_{ij}) \right\}.
\end{equation}
According to the conditional mean in \eqref{eq_mean}, the variance of \(Y_i\) is given by
\[
\text{Var}(Y_i) = 
\mu_i \left\{
\left( \frac{1 - \mu_i}{\mu_i} \right)^2
\exp\left( \frac{1 - \mu_i}{\mu_i} \right)
\, \text{Ei} \left(1, \frac{1 - \mu_i}{\mu_i} \right)
- \frac{1 - \mu_i}{\mu_i}
+ 1 \right\} - \mu_i^2,
\]
where \( \text{Ei}(a, z)\) is the exponential integral function. This model structure can be referred to as the UL-APLM. It allows for flexible modeling of additive terms through the smooth functions \(f_j(\cdot)\), while maintaining a linear term for the covariate vector \({\bf x}_i\).
		
\subsection{B-spline}
Each smooth function $f_j(t)$ is modeled using a B-spline basis of degree $k_j$, typically $k_j=3$ (cubic B-splines). Let \({\bf N}_j(t_{ij}) = \left\{ \text{N}_{j1}(t_{ij}), \ldots, \text{N}_{jq_j}(t_{ij}) \right\}^\T\) such that
$$
f_j(t_{ij}) = {\bf N}_j(t_{ij})^\T {\boldsymbol \gamma}_j,
$$
with ${\boldsymbol \gamma}_j = (\gamma_{j1}, \ldots, \gamma_{jq_j})^\T$ for $j=1, \ldots, r$. The ${\boldsymbol \gamma}_j$ are coefficients to be estimated.

\subsection{Penalized log-likelihood function}
The regular log-likelihood function for a single observation under the UL distribution is given by
\[
\ell_i(\mu_i) = 2 \log(1 - \mu_i) - \log(\mu_i) - 3 \log(1 - y_i) - \frac{y_i(1 - \mu_i)}{\mu_i (1 - y_i)}.
\]	
The total regular log-likelihood function is then given by \(\ell({\boldsymbol \theta}) = \sum_{i=1}^{n} \ell_i(\mu_i)\), where  ${\boldsymbol \theta} = ({\boldsymbol \beta}^\T, {\boldsymbol \gamma}_1^\T, \ldots, {\boldsymbol \gamma}_r^\T)^\T$, \(g(\mu_i) = {\bf x}_i^\T {\boldsymbol \beta} + \sum_{j=1}^{r} {\bf N}_j(t_{ij})^\T {\boldsymbol \gamma}_j \). To prevent overfitting caused by the potentially large number of spline coefficients, we control the roughness of the functions \( f_j(t) \) by incorporating a quadratic penalty that approximates the integral of their squared second derivatives. Specifically, we define the penalized log-likelihood function as
\begin{equation} \label{eq_penalty}
\ell_p({\boldsymbol \theta}, {\boldsymbol \alpha}) = \ell({\boldsymbol \theta}) - \frac{1}{2} \sum_{j=1}^{r} \alpha_j {\boldsymbol \gamma}_j^\T {\bf M}_j {\boldsymbol \gamma}_j,
\end{equation}
where \({\bf M}_j\) is a penalty matrix typically derived from the second-order difference matrix \({\bf D}_2\) as \({\bf M}_j = \{{\bf D}_2\}^\T {\bf D}_2 \) \citep{eilers_1996}. 
The smoothing parameters \( \alpha_j > 0 \) control the trade-off between fidelity to the data and smoothness of the estimated functions. 

The second-order difference matrix \( {\bf D}_2 \) is defined as the matrix of second-order finite differences of the B-spline basis functions. Given \( q \) as the number of basis functions for the spline, \( {\bf D}_2 \) is a \( (q-2) \times q \) matrix that computes the second differences between consecutive coefficients. Each entry in \( {\bf D}_2 \) corresponds to the difference between two adjacent spline coefficients, enforcing smoothness by penalizing large changes in the second derivative of the fitted curve. 

We adopt the \citet{wood_gam} basis reparameterization for model identifiability, with column centering and dropping the last column. The following notation continues under this reparameterization. The corresponding reparametrized second-order difference matrix is
\[
{\bf D}_2 = \begin{pmatrix}
-1 & 2 & -1 & 0 & \cdots & 0 \\
0 & -1 & 2 & -1 & \cdots & 0 \\
\vdots & \vdots & \vdots & \vdots & \ddots & \vdots \\
0 & \cdots & 0 & & 0 -1 & 2 \\
\end{pmatrix}.
\]
				
\subsection{Estimation}
MPLE of ${\boldsymbol \theta}$ is obtained maximizing \( \ell_p({\boldsymbol \theta}, {\boldsymbol \alpha}) \) for fixed values \( {\boldsymbol \alpha} = (\alpha_1, \ldots, \alpha_r)^\T \). The smoothing parameters may be chosen by methods such as cross-validation, generalized cross-validation (GCV), information criteria (AIC or BIC), or iterative updates \citep{wood_2017}.
		
The resulting model enables flexible, data-adaptive estimation of nonlinear relationships in the mean structure of the UL distribution, accommodating additive partially linear effects within a unified framework. To estimate the parameter vector ${\boldsymbol \theta}$, we maximize the penalized log-likelihood function defined in \eqref{eq_penalty}. This is achieved by setting the gradient (score function) of \( \ell_p({\boldsymbol \theta}, {\boldsymbol \alpha}) \) with respect to each parameter vector to zero. 

\subsubsection*{Penalized score function}
From supplementary \suppref{sec_1deriv} the penalized score functions for the parameters ${\boldsymbol \beta}$ and ${\boldsymbol \gamma}_j$ can be expressed, respectively, in matrix form as
\[
{\bf U}^{p}_{{\boldsymbol \beta}} = {\bf X}^\T {\bf W}^{(1)}{\bf 1}_n, \quad \text{and} \quad
{\bf U}^{p}_{{\boldsymbol \gamma}_j} = {\bf N}_j^\T {\bf W}^{(1)}{\bf 1}_n - \alpha_j {\bf M}_j {\boldsymbol \gamma}_j,
\]
where \({\bf X} \) is the $n \times p$ design matrix with rows ${\bf x}_i^\T$ , \({\bf N}_j \) is the $q_j \times n$ B-spline basis matrix with rows ${\bf N}_j(t_{ij})^\T$ corresponding to the covariate \( t_j \), \( {\bf W}^{(1)} \) is a diagonal matrix with elements 
\[
\text{W}_i^{(1)} = \left(\frac{-2}{1 - \mu_i} - \frac{1}{\mu_i} + \frac{y_i}{\mu_i^2 (1 - y_i)} \right)\mu_i (1 - \mu_i),
\]
and ${\bf 1}_n$ is a $n \times 1$ vector of ones. The MPLE \(\widehat{{\boldsymbol \theta}} \) is obtained by solving the system of equations ${\bf U}^{p}_{{\boldsymbol \beta}} = {\bf 0}$ and ${\bf U}^{p}_{{\boldsymbol \gamma}_j} = {\bf 0}$, for $j = 1, \ldots, r$. These equations are typically solved using a numerical optimization algorithm, such as Newton-Raphson or Fisher scoring, due to the nonlinearity of the penalized log-likelihood function and the inclusion of spline-based terms.

\subsubsection*{Penalized observed information matrix}
From supplementary \suppref{sec_2deriv} the penalized observed information matrix for 
${\boldsymbol \theta}$ is defined as
\[
\mathcal{I}_p^{({\boldsymbol \theta}, {\boldsymbol \theta})} =
\begin{bmatrix}
\mathcal{I}_p^{({\boldsymbol \beta}, {\boldsymbol \beta})} & \mathcal{I}_p^{({\boldsymbol \beta}, {\boldsymbol \gamma})} \\
\mathcal{I}_p^{({\boldsymbol \gamma}, {\boldsymbol \beta})} & \mathcal{I}_p^{({\boldsymbol \gamma}, {\boldsymbol \gamma})}
\end{bmatrix},
\]
where
$$
\mathcal{I}_p^{({\boldsymbol \beta}, {\boldsymbol \beta})}={\bf X}^\T {\bf W}^{(2)} {\bf X},
\quad 
\mathcal{I}_p^{({\boldsymbol \gamma}, {\boldsymbol \gamma})} =
{\bf N}^\T {\bf W}^{(2)} {\bf N} + \text{diag}\{\alpha_1{\bf M}_1, \ldots, \alpha_r {\bf M}_r\},
$$
and
$\mathcal{I}_p^{({\boldsymbol \beta}, {\boldsymbol \gamma})}=\{\mathcal{I}_p^{({\boldsymbol \gamma}, {\boldsymbol \beta})}\}^\T={\bf X}^\T {\bf W}^{(2)} {\bf N}$, with
${\bf N} = [{\bf N}_1, \ldots, {\bf N}_r]$ and ${\bf W}^{(2)}$ is a diagonal matrix with entries
\[
\begin{aligned}
\text{W}_i^{(2)} = & 
\left\{
\frac{2}{(1 - \mu_i)^2} - \frac{1}{\mu_i^2} + \frac{2 y_i}{\mu_i^3 (1 - y_i)}
\right\}
\left( \frac{1}{g'(\mu_i)} \right)^2 -
\text{W}_i^{(1)}
\frac{g''(\mu_i)}{\left\{g'(\mu_i)\right\}^3}.
\end{aligned}
\]

\subsubsection*{Iterative process} 
To estimate the parameter vector ${\boldsymbol \theta}$, we maximize the penalized log-likelihood function \eqref{eq_penalty}. Due to the nonlinearity of this function, the maximization problem does not admit closed-form solutions. We employ iterative algorithms to obtain the MPLEs, assuming that the smoothing parameters \({\boldsymbol \alpha} = (\alpha_1, \ldots, \alpha_r)^\T\) are fixed. Let us define the full design matrix as ${\bf N} = [{\bf X}, {\bf N}]$ 
and the block-diagonal penalty matrix as
$$
{\bf M}_\alpha = \text{blockdiag}\{{\bf 0}_{p}, \alpha_1 {\bf M}_1, \ldots, \alpha_r {\bf M}_r\},
$$
where {\bf 0}$_{p}$ is a $p \times p$ matrix of zeros and ${\bf M}_j$ is the penalty matrix corresponding to the smooth term ${\boldsymbol \gamma}_j$ for $j=1,\ldots,r$. Given the current ${\boldsymbol \theta}^{(m)}$ estimate, we compute the linear predictor
$$
\boldsymbol{\eta}^{(m)} = {\bf X} {\boldsymbol \beta}^{(m)} + \sum_{j=1}^r {\bf N}_j {\boldsymbol \gamma}_j^{(m)},
$$
and define the pseudo-response vector as
$$
\mathbf{z}^{(m)} = \boldsymbol{\eta}^{(m)} + \left({\bf W}^{(2,m)} \right)^{-1} \left({\bf W}^{(1,m)} \right){\bf 1}_n,
$$
where ${\bf W}^{(1,m)}$ and ${\bf W}^{(2,m)}$ are the diagonal matrices with entries $\text{W}_i^{(1)}$ and $\text{W}_i^{(2)}$ evaluated at ${\boldsymbol \theta}^{(m)}$. After some algebraic manipulation, one may show that the parameter vector updating has the following iteratively reweighted least squares (IRLS):
\begin{equation} \label{eq_irls}
\left( {\bf N}^\T {\bf W}^{(2,m)} {\bf N} + {\bf M}_\alpha \right) {\boldsymbol \theta}^{(m+1)} = {\bf N}^\T {\bf W}^{(2,m)} \mathbf{z}^{(m)}.
\end{equation}
Alternatively, a full update can be obtained using the penalized Fisher scoring matrix ${\bf K}_p^{({\boldsymbol \theta}, {\boldsymbol \theta})} = {\bf N}^\T {\bf P}^{(m)} {\bf N} + {\bf M}_\alpha$, with ${\bf P} = \text{diag}\{{\mathbb E}(\text{W}_1^{(2)}), \ldots, {\mathbb E}( \text{W}_n^{(2)})\}$ and ${\mathbb E}(\text{W}_i^{(2)})$ takes the same expression of $\text{W}_i^{(2)}$ with ${\mathbb E}\{Y_i/(1-Y_i)\} = \mu(1+\mu)/(1-\mu)$. The penalized Fisher scoring algorithm is also an IRLS of the form \eqref{eq_irls} changing ${\bf W}^{(2,m)}$ by ${\bf P}^{(m)}$.

The algorithm proceeds iteratively until the relative change in $\widehat {\boldsymbol \theta}$ falls below a predefined convergence threshold. This procedure is known as direct maximization or P-GAM type algorithm, unlike the traditional backfitting procedure \citep{stasinopoulos_gamlss} used in \texttt{gamlss} \citep{gamlss_2025} R package. Statistical inference for ${\boldsymbol \theta}$ may therefore based on the asymptotic normal distribution for $\widehat {\boldsymbol \theta}$ with variance-covariance matrix given by Var$(\widehat {\boldsymbol \theta}) = \{{\bf K}_p^ {({\boldsymbol \theta}, {\boldsymbol \theta})}\}^{-1}$.

\section{Diagnostic procedures}
\label{sec_diag}
Diagnostic tools are essential for evaluating the adequacy of the assumed model structure, 
including the correctness of the response distribution, the functional relationship between
the response and covariates, and the identification of influential or outlying observations. We present the quantile residuals  \citep{dunn_1996} adapted for the UL-APLM, along with a local influence approach \citep{cook_1986} to assess the sensitivity of parameter estimates 
under small perturbations in the data or model.

\subsection{Quantile residuals}
Quantile residuals are widely used in regression models to detect outliers and assess departures from the assumed distribution. They are particularly useful because they have an approximate standard normal distribution under correct model specification. For independent observations \( Y_i {\sim} \text{UL}(\mu_i) \), the quantile residual is defined as
\[
r_{q_i} = \Phi^{-1} \left\{ F_Y(y_i ; \widehat \mu_i) \right\}, \quad i = 1, \dots, n,
\]
where \( \Phi(\cdot) \) denotes the standard normal cumulative distribution function and \( F_Y(y; \mu) \) is the cumulative distribution function (CDF) of the unit-Lindley distribution given by
\[
F(y;\mu) = 1 - \left\{1 + \frac{(1 - \mu)y}{\mu(1 - y)} \right\} \exp \left\{-\frac{y(1 - \mu)}{\mu(1 - y)} \right\},
\quad 0 < y < 1,\ 0 < \mu < 1.
\]

Under the correctly specified model and for large \( n \), the quantile residuals \( r_{q_1}, \dots, r_{q_n} \) in the continuous case are approximately independent and identically distributed as standard normal. Therefore, plotting these residuals against the theoretical quantiles of the standard normal distribution in a Q-Q plot provides a visual diagnostic for assessing the goodness-of-fit of the fitted unit-Lindley regression model.

\subsection{Local influence analysis}
To investigate the sensitivity of the MPLEs in the UL-APLM, we adopt the local influence methodology introduced by \citet{cook_1986}, which has been extended to additive models. This technique evaluates the effect of small perturbations in the model or data through their impact on the penalized log-likelihood function. 

Let \( {\boldsymbol \omega} = (\omega_1, \ldots, \omega_n)^\T \) be a vector of perturbation parameters. The perturbed penalized log-likelihood function is denoted by $\ell_p({\boldsymbol \theta}, {\boldsymbol \alpha} \mid {\boldsymbol \omega})$, such that
\[
\ell_p({\boldsymbol \theta}, {\boldsymbol \alpha} \mid {\boldsymbol \omega}_0) = \ell_p({\boldsymbol \theta}, {\boldsymbol \alpha}),
\]
where ${\boldsymbol \omega}_0$ represents the vector of unperturbation parameters. The likelihood displacement function is defined as 
\[
\text{LD}({\boldsymbol \omega}) = 2\left\{ \ell_p(\widehat{{\boldsymbol \theta}}, {\boldsymbol \alpha} \mid {\boldsymbol \omega}_0) - 
\ell_p(\widehat{{\boldsymbol \theta}}_{{\boldsymbol \omega}}, {\boldsymbol \alpha} \mid {\boldsymbol \omega}) \right\},
\]
where \( \widehat{{\boldsymbol \theta}} \) and \( \widehat{{\boldsymbol \theta}}_{{\boldsymbol \omega}} \) 
are, respectively, the MPLEs under unperturbed and perturbed settings. The normal curvature of \( \text{LD}({\boldsymbol \omega}) \) in the direction \( {\bf D} \in \mathbb{R}^n \), \( \|{\bf D}\| = 1 \), defined as
\[
C_{{\bf D}}(\widehat{{\boldsymbol \theta}}) = 2 \left| {\bf D}^\T {\boldsymbol \Delta}^\T 
\left\{\widehat{\mathcal{I}}_p^{({\boldsymbol \theta}, {\boldsymbol \theta})}\right\}^{-1} {\boldsymbol \Delta} {\bf D} \right|,
\]
where $\widehat{\mathcal{I}}_p^{({\boldsymbol \theta}, {\boldsymbol \theta})}$ is the penalized observed information matrix
evaluated at $\widehat {\boldsymbol \theta}$ and  
\[
{\boldsymbol \Delta} = \left. \frac{\partial^2 \ell_p({\boldsymbol \theta}, {\boldsymbol \alpha} \mid {\boldsymbol \omega})}{\partial {\boldsymbol \theta} \partial {\boldsymbol \omega}^\T} \right|_{\hat{{\boldsymbol \theta}}, {\boldsymbol \omega}_0}
\]
is the sensitivity matrix with respect to the perturbation scheme. 

The most influential direction, \( |{\bf D}_{\max}| \), is the eigenvector corresponding to the largest eigenvalue of the matrix  
\[
{\bf B} = {\boldsymbol \Delta}^\T \left\{\widehat{\mathcal{I}}_p^{({\boldsymbol \theta}, {\boldsymbol \theta})}\right\}^{-1}  {\boldsymbol \Delta}.
\]

A scale-invariant measure of influence, the conformal normal curvature \citep{poon_1999} with values in \( [0,1] \), is computed as
$$
B_{{\bf D}}(\hat{{\boldsymbol \theta}}) = 
\frac{\left|{\bf D}^\T {\bf B} {\bf D}\right|}{
\sqrt{\operatorname{tr}\left( {\bf B}^2\right)}}.
$$

To identify influential observations, two diagnostic measures are proposed. The first is the directional curvature for the \(i\)th observation, given by
\[
m_i^{(q)} = \sqrt{\sum_{j=1}^k \widehat{\lambda}_j e_{ji}^2},
\]
where \(\widehat{\lambda}_j \) are the normalized eigenvalues and \( e_{ji} \) the corresponding components of the eigenvectors of the matrix ${\bf B}/\sqrt{\operatorname{tr}\left( {\bf B}^2\right)}$. The cutoff value \( k \) is determined by eigenvalues satisfying \(\widehat{\lambda}_j > q / \sqrt{n} \). The second measure is conformal curvature in the direction of the canonical basis vector, defined as
\[
\text{B}_i = \left\{ m_i^{(0)} \right\}^2 = \sum_{j=1}^n \widehat{\lambda}_j e_{ji}^2.
\]
An observation is flagged as influential if \( \text{B}_i > \bar{\text{B}} + c \cdot sd(\text{B}) \), 
where $sd(\text{B})$ denotes the standard deviation of $\text{B}_1, \ldots, \text{B}_n$ and \( c \) as a user-defined constant. If interest lies in the local influence on a specific subset of parameters (such as the linear coefficients \( {\boldsymbol \beta} \) or a particular smooth term coefficients \( {\boldsymbol \gamma}_j \)), the inverse penalized observed information matrix can be restricted to the relevant block to compute corresponding curvatures. 

Together, the inverse-penalized observed information and the sensitivity matrix enable the computation of local influence measures, such as normal curvature and conformal normal curvature, to identify influential data points. In the following, we explicitly derive the components of the sensitivity matrix for two common perturbation schemes.

\subsubsection*{Case-weight perturbation}
In this perturbation scheme for $\omega_i \in [0,1]$, the perturbed penalized log-likelihood function is
\[
\ell_p({\boldsymbol \theta}, {\boldsymbol \alpha} \mid {\boldsymbol \omega}) = \sum_{i=1}^n \omega_i \ell_i(\mu_i) - \frac{1}{2} \sum_{j=1}^r \alpha_j {\boldsymbol \gamma}_j^\T {\bf M}_j {\boldsymbol \gamma}_j,
\]
and the reference point is ${\boldsymbol \omega}_0 = \mathbf{1}$. The sensitivity matrix is
\[
{\boldsymbol \Delta} = \left. {\bf N}^\T {\bf W}^{(1)}{\bf I}_n \right|_{\hat{{\boldsymbol \theta}}, {\boldsymbol \omega}_0} = {\bf N}^\T \hat{{\bf W}}^{(1)}.
\]

\subsubsection*{Response perturbation}
In this perturbation scheme, the observed response is perturbed as $y_{i\omega} = y_i + \omega_i$, where $\omega_i \in \mathbb{R}$ and ${\boldsymbol \omega}_0 = {\bf 0}$. Consequently, the perturbed penalized log-likelihood function depends on $\omega_i$ only through $y_{i\omega}$. The sensitivity matrix is
\[
{\boldsymbol \Delta} = \left. {\bf N}^\T \text{diag}\left\{ \frac{\mu_i (1 - \mu_i)}{\mu_i^2} \frac{\partial}{\partial \omega_i} \left(\frac{y_{i\omega}}{ 1 - y_{i\omega}} \right)\right\} \right|_{\hat{{\boldsymbol \theta}}, {\boldsymbol \omega}_0} = {\bf N}^\T \hat {\bf W}^{(3)},
\]
in which ${\bf W}^{(3)} = \text{diag}\{\text{W}^{(3)}_1, \cdots, \text{W}^{(3)}_n\}$ with
\[
\text{W}^{(3)}_i = \frac{\mu_i (1 - \mu_i)}{\mu_i^2 (1 - y_i)^2}.
\]

\subsection{Effective degrees of freedom}
In the context of the UL-APLM, we define the effective degrees of freedom (\textit{edf}) as the complexity of the fitted linear predictor $\widehat {\boldsymbol \eta} = {\bf N} \widehat{\boldsymbol \theta}$. At convergence, the IRLS algorithm used to obtain $\widehat{{\boldsymbol \theta}}$ yields the closed-form solution in \eqref{eq_irls}. The smoother matrix, which projects the working response on the linear predictor space, is given by
$$
\widehat{\bf H}_\alpha = {\bf N}\left({\bf N}^\T \widehat{{\bf W}}^{(2)} {\bf N} + {\bf M}_\alpha \right)^{-1} {\bf N}^\T \widehat{{\bf W}}^{(2)},
$$
and the trace of the smoother matrix provides the effective degrees of freedom
$$
\textit{edf}_\alpha =  \text{tr} \left\{ \left({\bf N}^\T \widehat{{\bf W}}^{(2)} {\bf N} + {\bf M}_\alpha \right)^{-1} 
{\bf N}^\T \widehat{{\bf W}}^{(2)}{\bf N} \right\}.
$$

From \citet{eilers_1996} we may establish the following relationship:
\begin{align*}
\text{tr} (\widehat{{\bf H}}_\alpha) & = \text{tr} \left\{ \left( \mathbf{Q} + {\bf M}_\alpha \right)^{-1} \mathbf{Q} \right\} \\
& = \text{tr} \left\{ \mathbf{Q}^{1/2} \left( \mathbf{Q} + {\bf M}_\alpha \right)^{-1} \mathbf{Q}^{1/2} \right\} \\
& = \text{tr} \left\{ \left( \mathbf{I}_{p + q_1 + \ldots + q_r} + \mathbf{L} \right)^{-1} \right\} \\
& = p + \sum_{l=1}^{q_1 + \ldots + q_r} \frac{1}{1+\lambda_l({\boldsymbol \alpha})},\\ 
\end{align*}
where $\mathbf{Q} = {\bf N}^\T \widehat{{\bf W}}^{(2)} {\bf N}$ and $\lambda_l({\boldsymbol \alpha})$ are the eigenvalues of the symmetric nonnegative-definite matrix $\mathbf{L} = \mathbf{Q}^{-1/2} {\bf M} \mathbf{Q}^{-1/2}$. So, one has that $ p < \textit{edf}_\alpha < p + q_1 + \ldots + q_r$ and specifically the effective degrees of freedom for $\widehat{\boldsymbol \gamma}_1$ correspond to the sum of the principal diagonal elements of the matrix $\left( \mathbf{I}_{p + q_1 + \ldots + q_r} + \mathbf{L} \right)^{-1}$ in the positions $p+1$ to $p+q_1$. The same reasoning applies to the estimates $\widehat{\boldsymbol \gamma}_2, \ldots, \widehat{\boldsymbol \gamma}_r$.

\subsection{Smoothing parameter estimation}
As highlighted by \citet{stasinopoulos_gamlss}, several methods exist for selecting the smoothing parameters ${\boldsymbol \alpha}$. For example, as global methods, we have the maximization of the marginal log-likelihood function $\ell_p({\boldsymbol \alpha})$, the minimization of the Akaike criterion $\text{AIC}_\alpha = -2 \ell_p({\boldsymbol \theta}, {\boldsymbol \alpha}) + 2 \textit{edf}_\alpha$ and the generalized cross-validation \citep{craven_1979}, which consists in minimizing the quantity 
\[
\text{GCV}_\alpha = \frac{n\sum_{i=1}^n \left( y_i - \hat{\mu}_i \right)^2}{\{n - \textit{edf}_\alpha \}^2}.
\]
There are also local methods based on partial residuals from the backfitting procedure. 

We employed the global generalized Fellner-Schall method \citep{wood_2017}. This method provides an explicit update for smoothing parameters, which control the trade-off between model fit and smoothness of the estimated functions \( f_j(t) \). Let ${\bf M}_\alpha^{-}$ the Moore-Penrose pseudo-inverse of ${\bf M}_\alpha$. Then the Laplace approximation of the marginal log-likelihood is
\begin{align*}
\ell_p({\boldsymbol \alpha}) & = \ell({\boldsymbol \theta}) - \frac{1}{2}\log{ \Big| {\bf M}_\alpha^{-} \Big| } - \frac{1}{2} {\boldsymbol \theta} ^\T {\bf M}_{\alpha} {\boldsymbol \theta} + \frac{1}{2}\log{\Big| \left\{\widehat{\mathcal{I}}_p^{({\boldsymbol \theta}, {\boldsymbol \theta})}\right\}^{-1} \Big|} + \text{const}.
\end{align*}
For $\widetilde{{\bf M}}_j = \text{blockdiag}\{{\bf 0}_{p}, {\bf 0}_{q_1}, \ldots, \alpha_j {\bf M}_j, \ldots, {\bf 0}_{q_r}\}$, the derivative of the penalized log-likelihood with respect to $\alpha_j$ is expressed as 
\begin{align*}
\frac{\partial \ell_p({\boldsymbol \alpha})}{\partial \alpha_j} & = \frac{1}{2} \text{tr} \left\{ {\bf M}_\alpha^{-} \widetilde{\bf M}_j \right\} - \frac{1}{2} {\boldsymbol \theta} ^\T \widetilde{\bf M}_j {\boldsymbol \theta} - \frac{1}{2} \text{tr} \left\{ \left\{\widehat{\mathcal{I}}_p^{({\boldsymbol \theta}, {\boldsymbol \theta})}\right\}^{-1} \widetilde{\bf M}_j \right\}.
\end{align*}

An efficient and simple update formula for smoothing optimization via the generalized Fellner-Schall method for UL-APLM, which guarantees an increase in the marginal log-likelihood at each iteration, is given by
\begin{align*}
\alpha_j^{(m+1)} & = \frac{\text{tr} \left\{ {\bf M}_\alpha^{-} \widetilde{\bf M}_j \right\} - \text{tr} \left\{ \left\{\widehat{\mathcal{I}}_p^{({\boldsymbol \theta}, {\boldsymbol \theta})}\right\}^{-1} \widetilde{\bf M}_j \right\}}{ \left\{ {\boldsymbol \gamma}_j^{(m)} \right\} ^\T {\bf M}_j {\boldsymbol \gamma}_j^{(m)} } \alpha_j^{(m)}.
\end{align*}
A major advantage of this method lies in its efficiency and computational simplicity. We alternated smoothing-parameter updates at each step of the algorithm in \eqref{eq_irls}, continuing until convergence of the model coefficients is achieved.

\section{Simulation studies} \label{sec_simu}
To evaluate the finite-sample performance of the proposed UL-APLM, which models additive terms via penalized B-splines, a Monte Carlo simulation study was conducted. The main objectives of this study were to assess the bias and consistency of the linear coefficient estimates and evaluate the accuracy of the estimated additive functions.

\begin{figure}[htb]
\centering
\includegraphics[width=13cm]{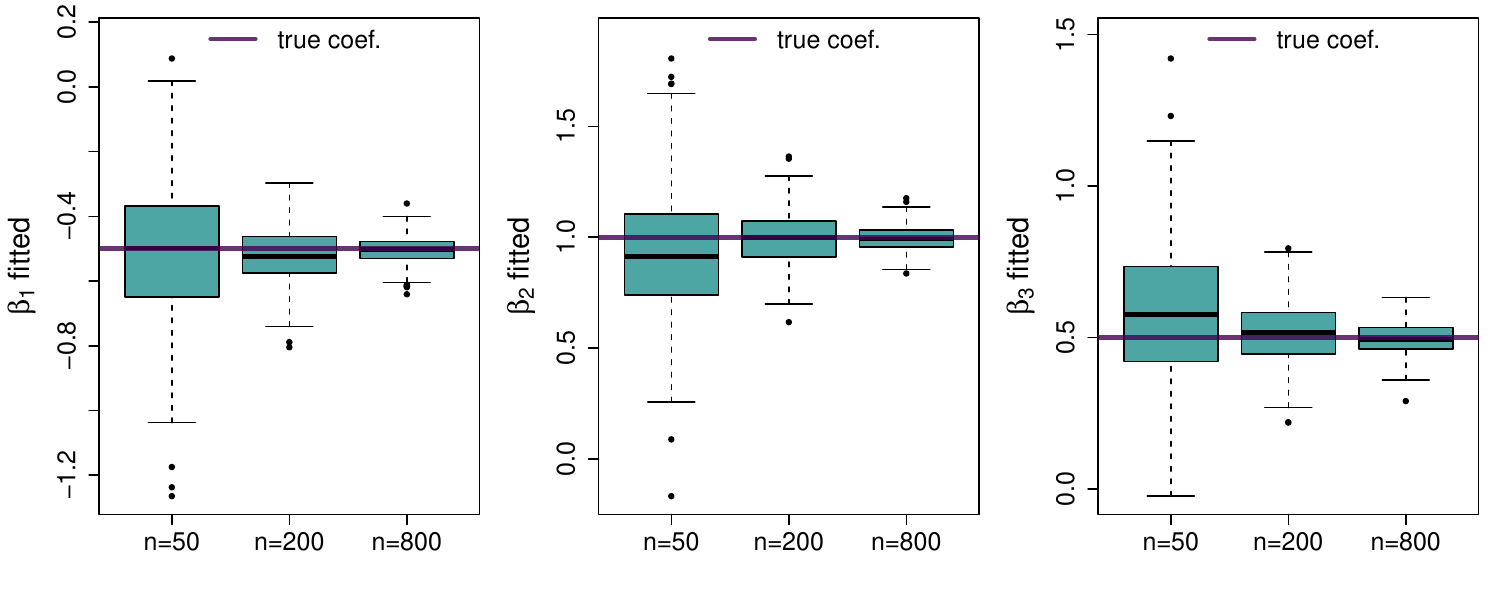}
\includegraphics[width=13cm]{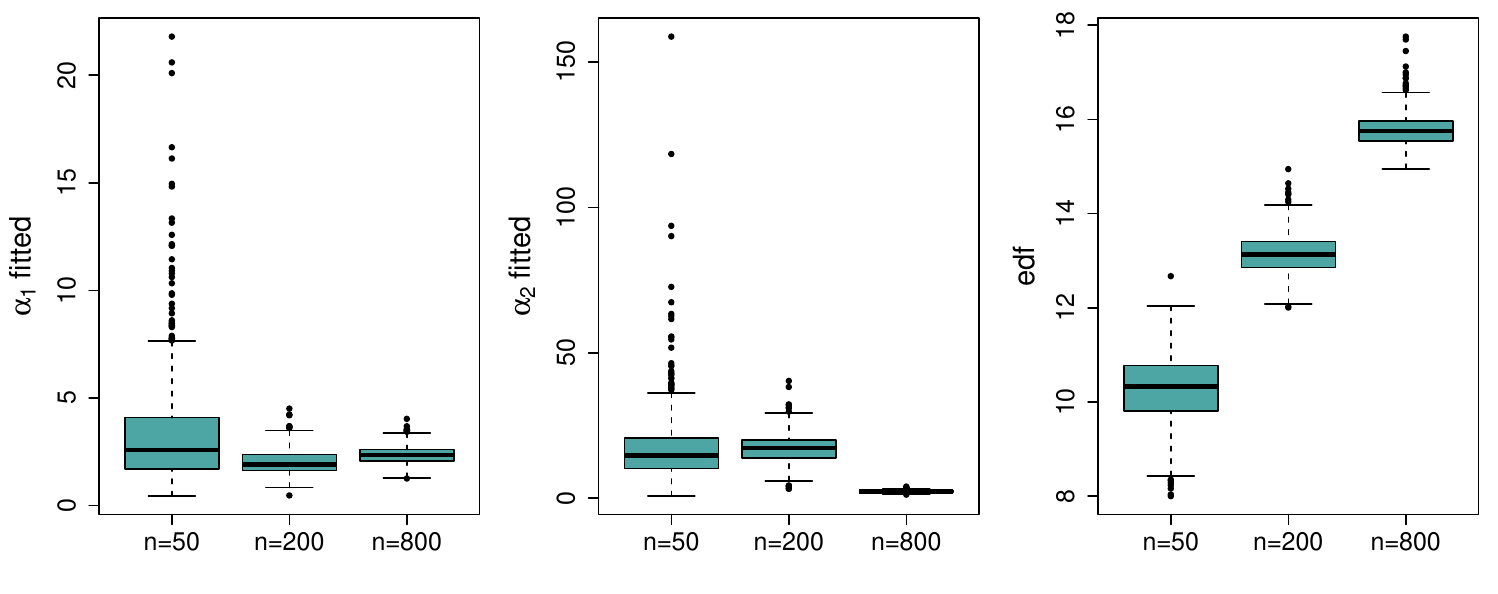}
\caption{Simulation results based on 500 replications for sample sizes $n = 50, 200, 800$ under the proposed UL-APLM: Fitted linear part coefficients (top), smoothing parameter estimates, and computed \textit{edf} (bottom).}
\label{simu_coefs}
\end{figure}
The simulation was replicated $\text{R} = 500$ times for sample sizes $n = 50, 200, 800$. True parameter vector for the linear part was set as $\beta = (-0.5, 1, 0.5)^\T$. For each replication, the following steps were performed:
\begin{itemize}
\item[-] Two covariates were generated and fixed: a binary variable $x_1 \sim \text{Bernoulli}(0.5)$ and a continuous variable $x_2 \sim \text{Uniform}(-1, 1)$. Two independent additive terms were constructed based on continuous variables $t_1 \sim \text{Uniform}(-5, 0)$ and $t_2 \sim \text{Uniform}(0, 3)$. The associated true smooth functions were defined as centered versions of
$$  
f_1(t_1) = \cos(0.5\pi t_1) \quad \quad \text{and} \quad \quad  f_2(t_2) = \frac{-t_2^3 + \exp(t_2)/4}{5}.
$$

\item[-] The linear predictor was computed as $\eta_i$, and the conditional mean was obtained using the inverse logit function $\mu_i = (1 + \exp(-\eta_i))^{-1}$. The response variables  \( Y_i \ {\sim} \ \text{UL}(\mu_i) \) were then generated from the unit-Lindley distribution with mean $\mu_i$ (see \suppref{sec_Rgen} for the data-generation R function).

\item[-] The model coefficients and smooth parameters were estimated with the procedure developed in this text, with both additive terms represented by cubic B-splines with second-order penalization and 8 clamped knots equally spaced. The MPLEs $\hat \beta_1$, $\hat \beta_2$ and $\hat \beta_3$ were stored as well as the fitted values $\hat{f}_1(t_1)$ and $\hat{f}_2(t_2)$. We used $\alpha_j^{(0)} \sim \text{Uniform}(1, 1000)$ as initial values in each replicate to check the robustness of the smoothing procedure estimation (see \suppref{sec_Rulaplm} for the model-fitting R function).

\item[-] After 500 replications, the boxplot for each estimated coefficient and smooth parameter was computed. Additionally, the fitted functions $\hat{f}_1(t_1)$ and $\hat{f}_2(t_2)$ were plotted against their true counterparts to visualize the adequacy and variability of smooth function recovery (see \suppref{sec_Rsimu} for simulation R commands).
\end{itemize}

\begin{figure}[htb]
\centering
\includegraphics[width=13cm]{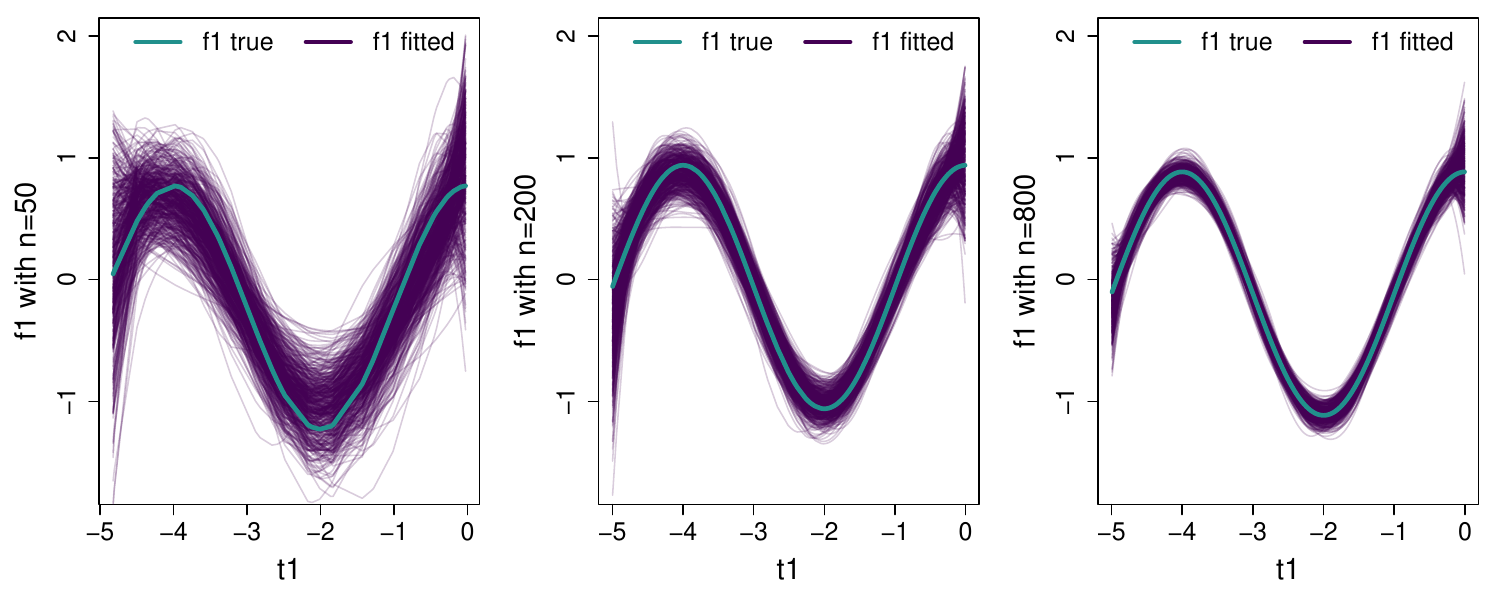}
\includegraphics[width=13cm]{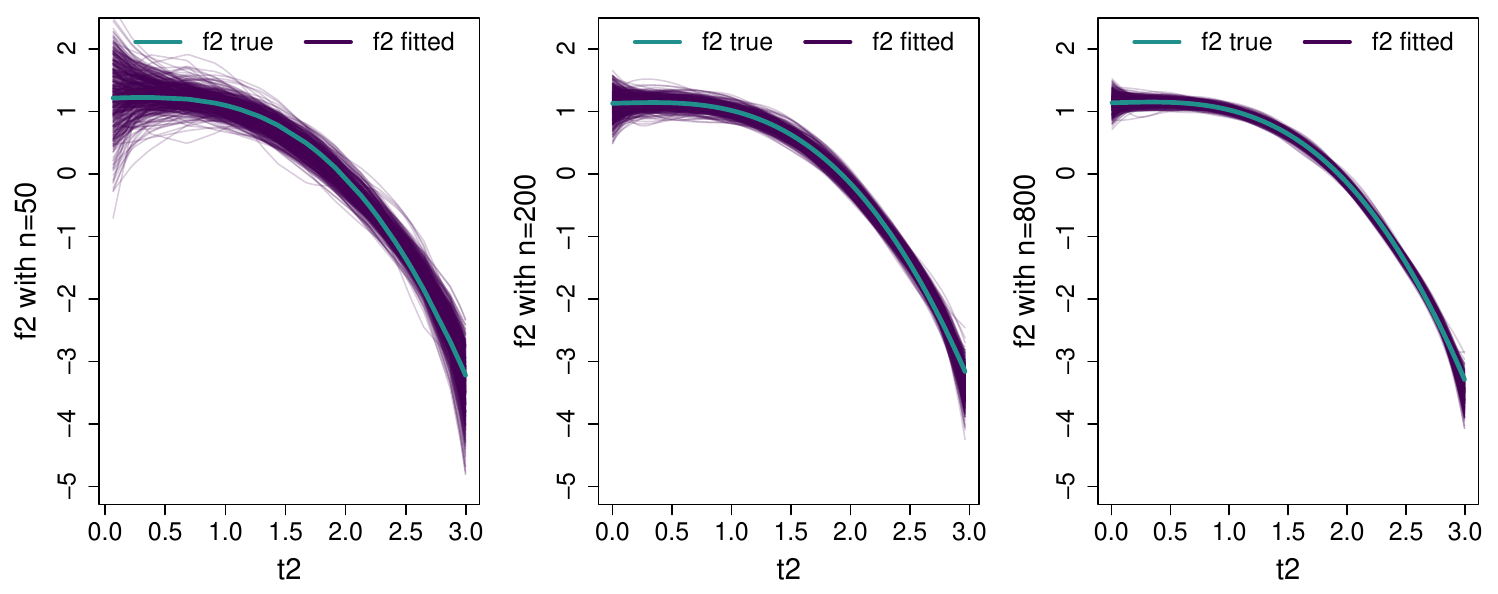}
\caption{Simulation results based on 500 replications for sample sizes $n = 50, 200, 800$ under the proposed UL-APLM: Estimated versus true functions $f_1(t_1)$ (top) and $f_2(t_2)$ (bottom).}
\label{simu_f}
\end{figure}
\autoref{simu_coefs} presents the simulation results based on 500 replications for three different sample sizes, $n = 50, 200, 800$, under the proposed UL-APLM. We can see that the estimation procedure is consistent in $n$, works even for small sample sizes, and is robust to initial values of the smoothing parameter. The simulation procedure was computationally efficient, requiring approximately 110 seconds to generate the data and perform all models on a personal notebook with a 13th Gen Intel(R) Core(TM) i5-13450HX (8x 2.40 GHz) processor and 16.0 GB of RAM. The estimation algorithm demonstrated numerical stability, and all models converged satisfactorily.

Overall, the results in \autoref{simu_f} confirm that the UL-APLM provides accurate and reliable estimates, particularly when a sufficient sample size is available. The visual reduction in both bias and MSE with larger $n$ highlights the robustness and consistency of the estimation approach.

\section{Application} \label{sec_aplic}
To illustrate the proposed methodology, we analyzed a dataset from the Center of Applied Statistics of the University of São Paulo \citep{paula_2023}, which concerns the psychological profiles of patients with hypopituitarism (see \suppref{sec_dataset}). The sample comprises $n=54$ patients aged 25-60 years. The following variables were considered: (i) {\tt resp}: emotional control (standardized on the unit interval), (ii) {\tt dep}: depression (standardized on the unit interval), (iii) {\tt adreno}: deficiency of the adrenocorticotropic hormone ({\tt no}, {\tt yes}), (iv) {\tt drug1}: use of growth hormone ({\tt no}, {\tt yes}) and {\tt drug2}: use of only prednisone or hydrocortisone ({\tt none}, {\tt predi}, {\tt hicro}). All information refers to the 12 months before the medical evaluation.

\begin{figure}[htb]
\centering
\includegraphics[width=13cm]{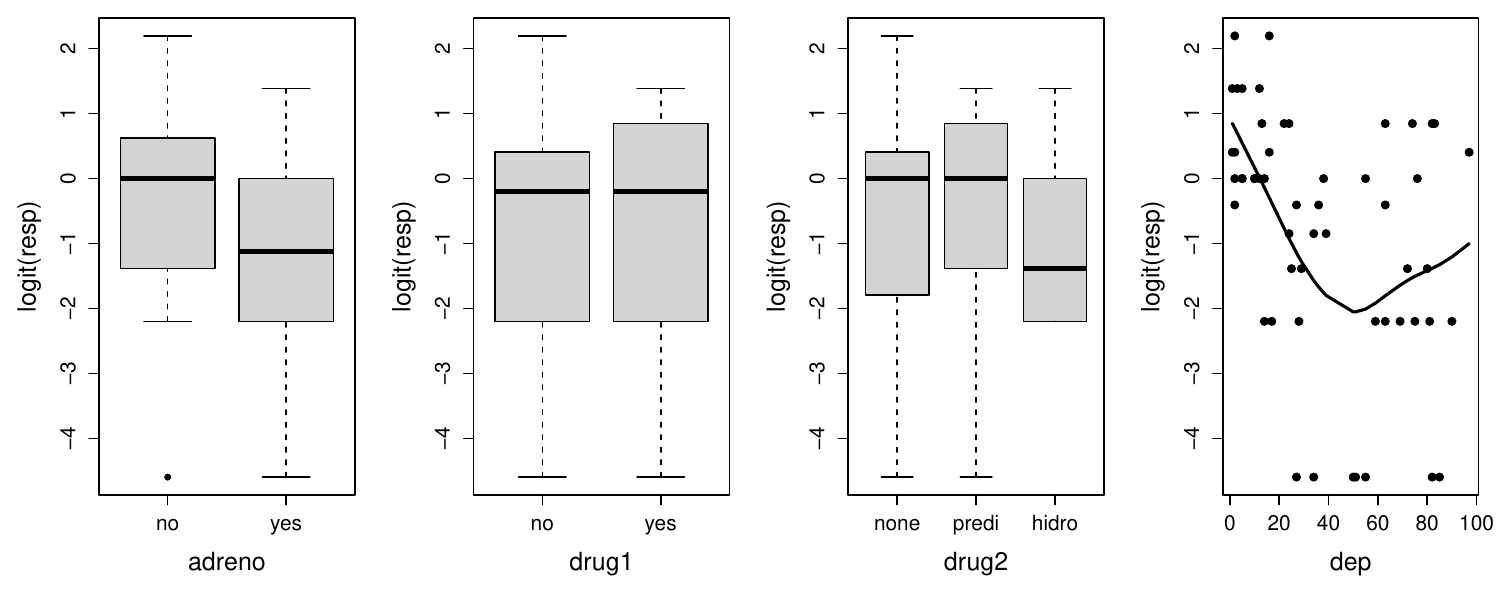}
\caption{Descriptive analysis of the dataset on hypopituitarism.}
\label{hypo_desc}
\end{figure}
\begin{table}[htb]
\centering
\caption{Coefficient estimates and approximate standard errors from the fitted UL-APLM with all observations.}
\begin{tabular}{crrrr}
\toprule
Parameter   & Estimate  & Std. Error & Z-value &  P-value\\
\midrule
$\beta_1$  &  -0.10 & 0.18  & -0.54  & 0.294\\
$\beta_2$  &  -0.44 & 0.25  & -1.79  & 0.037\\
$\beta_3$  &  0.67  &  0.43  & 1.56  & 0.059\\
$\beta_4$   &   -0.53 &  0.45  &  -1.18 & 0.119\\
$\beta_5$   &   -1.02 &  0.42  &  -2.46  & 0.007\\
\midrule
\textit{edf} & 7.797 & - & - & -\\
$\alpha$ & 21.082 & - & - & -\\
\botrule
\end{tabular}
\label{hypo_b1_coefs}
\end{table}

\begin{figure}[htb]
\centering
\includegraphics[width=13cm]{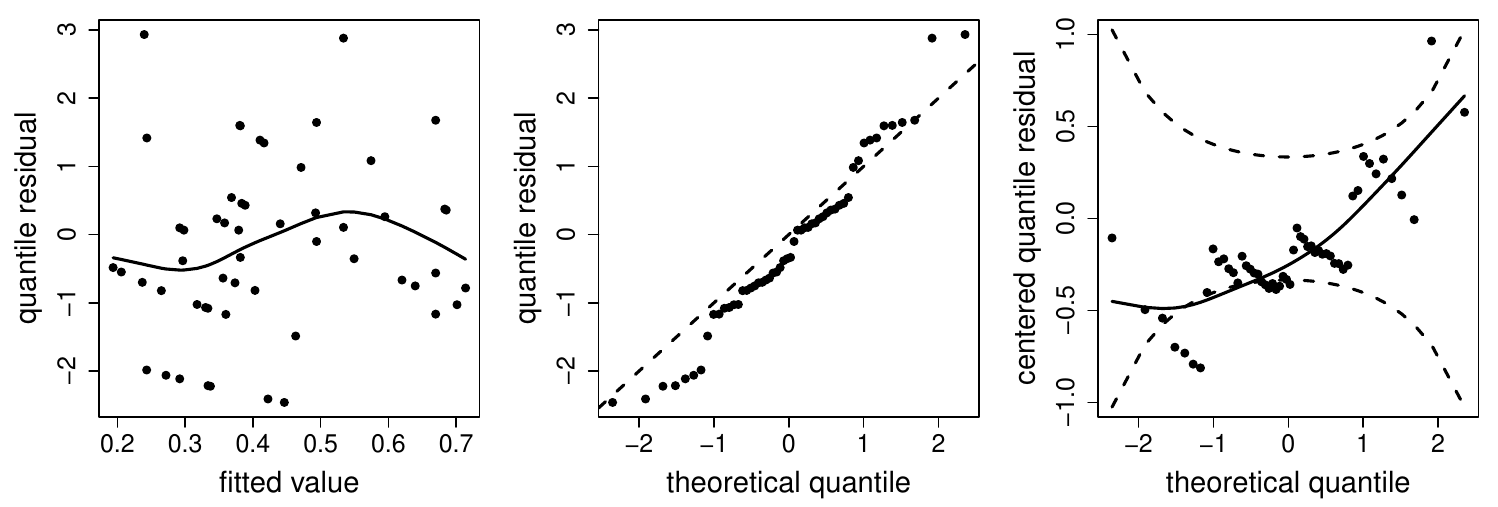}
\includegraphics[width=13cm]{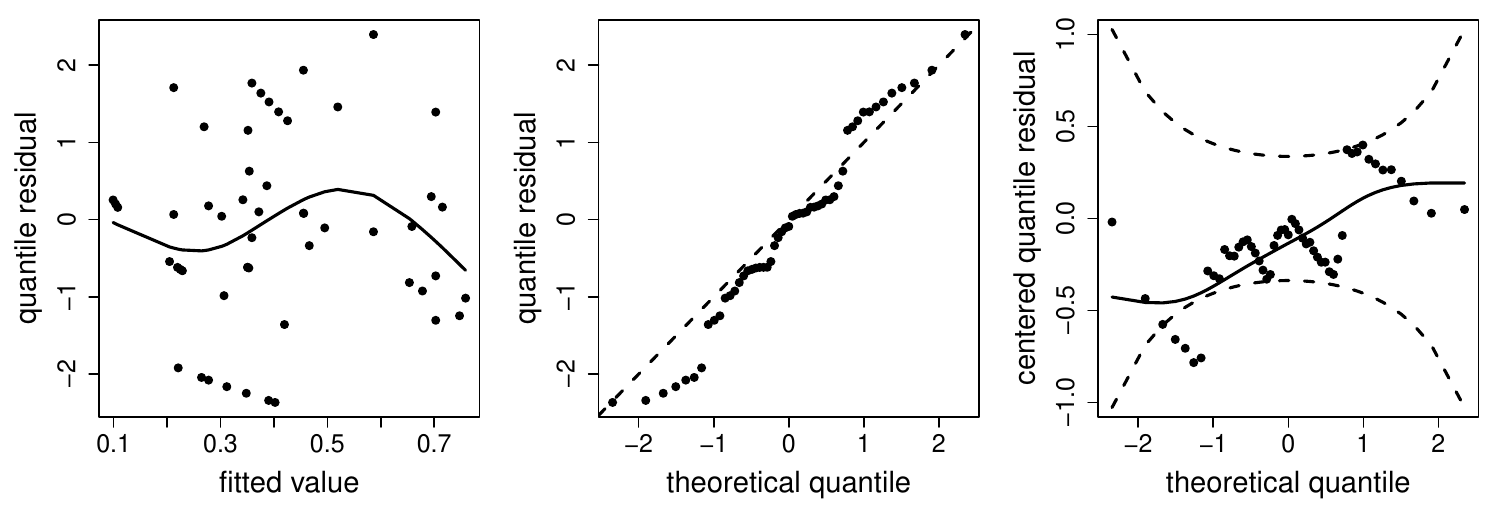}
\includegraphics[width=13cm]{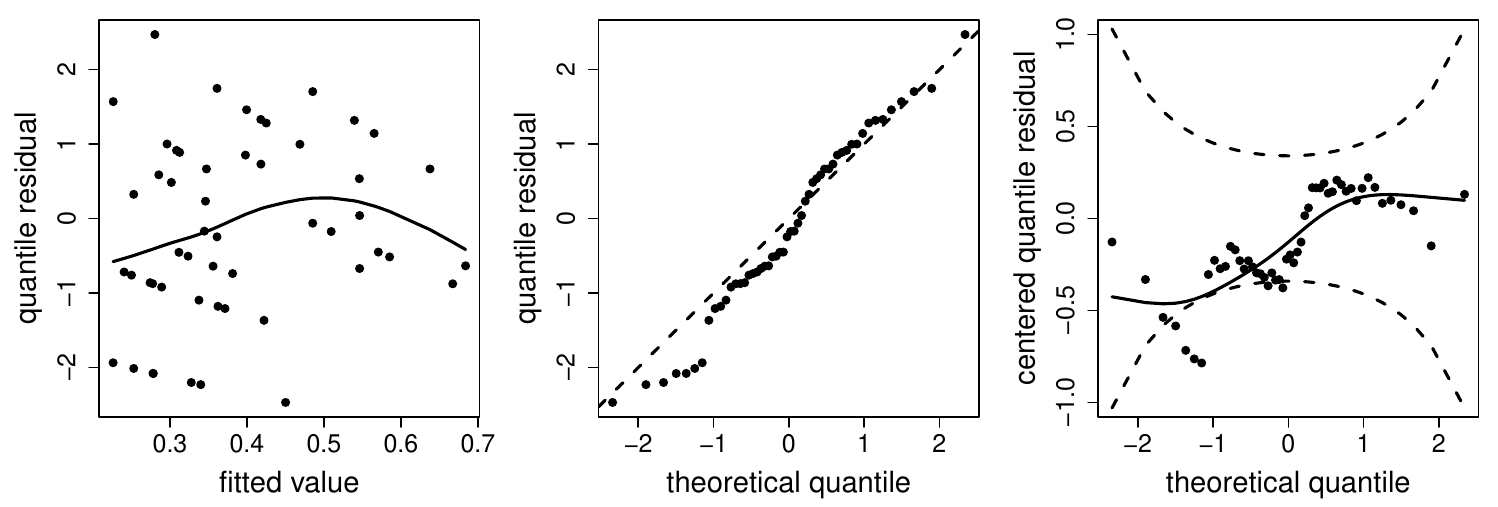}
\caption{Quantile residual versus fitted value (left), quantile residual (center) and centered quantile residual (right) versus theoretical quantiles of the standard normal distribution for the fitted model with all observations (top), dropping observation \textbf{11} (middle) and dropping observations \textbf{43} and \textbf{47} (bottom).}
\label{hypo_b}
\end{figure}

A descriptive data analysis of the data is shown in \autoref{hypo_desc}. From this figure, a nonlinear relationship between the response variable (emotional control) and depression is evident. In addition, the boxplots indicate that patients with adrenocorticotropic hormone deficiency generally have lower emotional control, whereas those using growth hormone or only prednisone tend to have higher emotional control. Based on these patterns, we propose the following UL-APLM to model the mean response:
\begin{itemize}
\item[-] $\texttt{resp}_i|{\bf x}_i \stackrel{\rm ind} {\sim} \text{UL}(\mu_i)$
\item[-] $\text{logit}\left (\mu_i \right) = \beta_1 + \beta_2 \texttt{adreno}_i + \beta_3 \texttt{drug1}_i + \beta_4 \texttt{predi}_i + \beta_5 \texttt{hicro}_i + f(\texttt{dep}_i),$
\end{itemize}
where ${\bf x}_i = (\texttt{adreno}_i, \texttt{drug1}_i, \texttt{predi}_i, \texttt{hicro}_i, \texttt{dep}_i)^\T$, with $\texttt{predi}_i$ and $\texttt{hicro}_i$ being binary variables corresponding to $\texttt{drug}_2$, and $f(\texttt{dep}_i)$ is a smooth function modeled by a cubic B-splines with second-order penalization and eight equally spaced clamped knots (see \suppref{sec_Rhypo} for R commands related to this application). 

The coefficient estimates in \autoref{hypo_b1_coefs} indicate that $\beta_2, \beta_3, \beta_5$ from linear part effects are marginally significant. For the additive term, the estimated coefficients are $\widehat \gamma_1$ = 0.80  (0.63), $\widehat \gamma_2$ = 0.57 \ (0.59), $\widehat \gamma_3$ = 0.31 \ (0.59), $\widehat \gamma_4$ = -0.04 \ (0.60), $\widehat \gamma_5$ =  -0.41 \ (0.62), $\widehat \gamma_6$ =  -0.64 \ (0.65), $\widehat \gamma_7$ =  -0.69 \ (0.65), $\widehat \gamma_8$ =  -0.59 \ (0.64), $\widehat \gamma_9$ =  -0.43 \ (0.58), $\widehat \gamma_{10}$ =  -0.27 \ (0.47) and $\widehat \gamma_{11}$ =  -0.13 \ (0.28), with $\textit{edf}_\gamma = 7.797 - 5 = 2.797$. The joint Wald test to assess $\text{H}_0: \gamma_1 = \ldots = \gamma_{11} = 0$ yields a P-value $< 0.001$, indicating that there is some statistically significant nonlinear tendency to explain the expected emotional control by depression. 
\begin{figure}[htb]
\centering
\includegraphics[width=13cm]{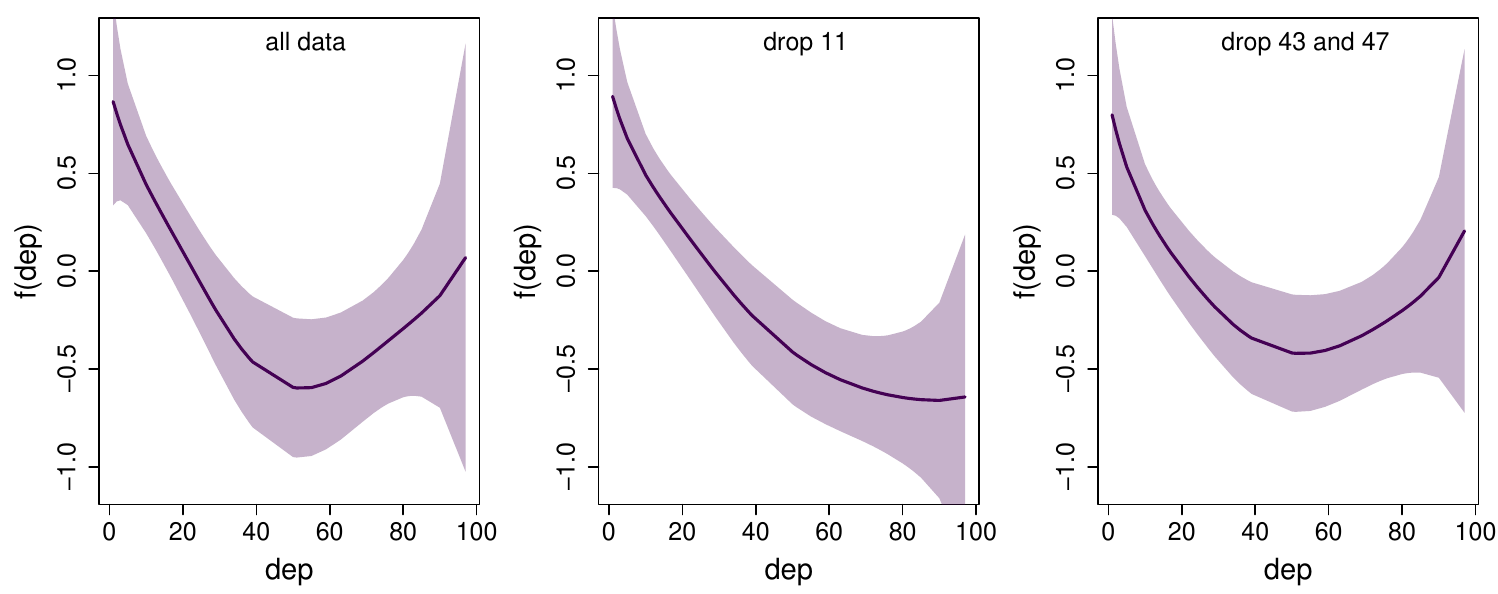}
\caption{Fitted additive function for depression with all observations (left), dropping observation \textbf{11} (center), and dropping observations \textbf{43} and \textbf{47} (right).}
\label{hypo_f}
\end{figure}
\begin{figure}[htb]
\centering
\includegraphics[width=13cm]{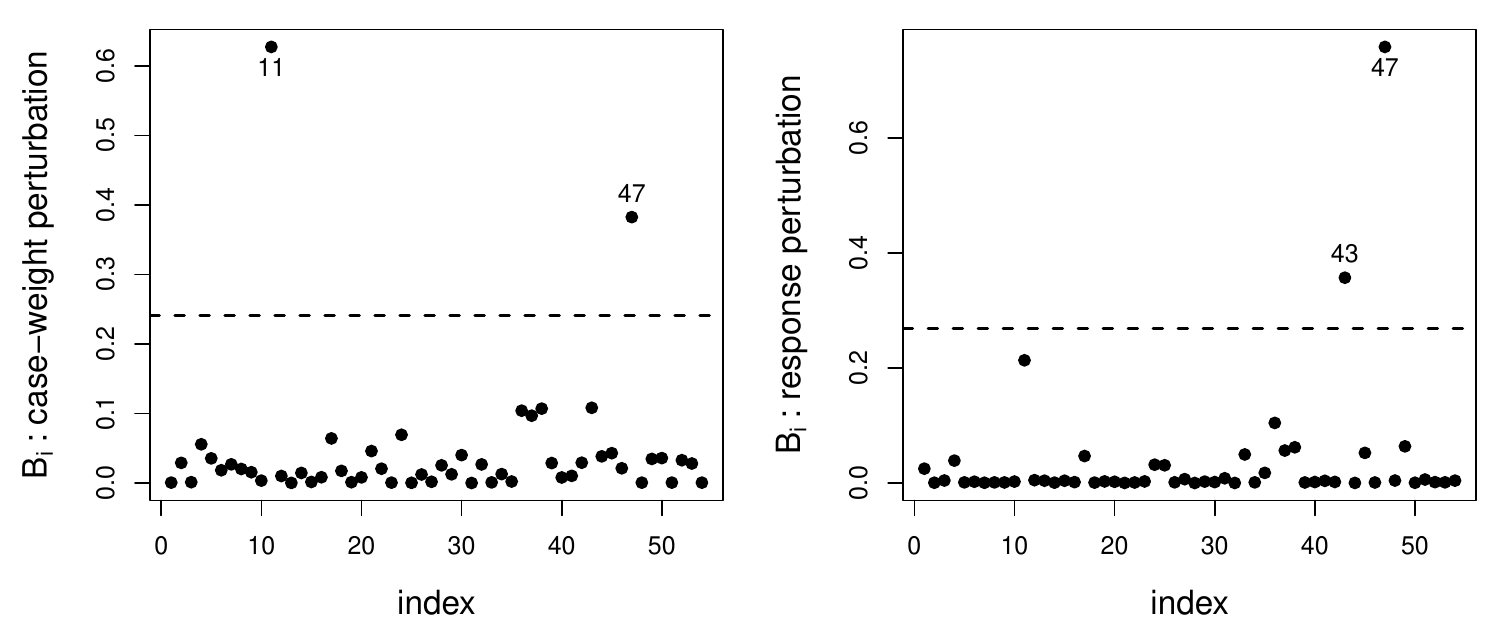}
\caption{Influence measure B$_i$ versus observation index of the fitted model with all observations under case-weight and response perturbation schemes.}
\label{hypo_influ}
\end{figure}
Patients with adrenocorticotropic hormone deficiency and those using only prednisone or only hydrocortisone exhibit lower emotional control, whereas patients treated with growth hormone show higher emotional control. From the residual analysis in \autoref{hypo_b}, the proposed UL-APLM seems to be suitable for this dataset, and the tendency in \autoref{hypo_f} (left) closely matches the descriptive analysis.

The local influence plots shown in \autoref{hypo_influ} (see \suppref{sec_Rdiag} for residual and local influence tools in R) depict the influence measures computed directly from the fitted model. In \autoref{hypo_f} (center and right), the effects of the influential observations detected in the local influence graphs on the estimated $f(\texttt{dep}_i)$ are compared. Observations \textbf{11}, \textbf{43}, and \textbf{47} are highlighted in influence graphs under case-weight and response perturbation schemes. Observation \textbf{11} alters the final tendency of $f(\texttt{dep}_i)$, whereas removing observations (\textbf{43}, \textbf{47}) improves the goodness-of-fit. Patient \textbf{11} exhibits an atypical profile-high emotional control combined with high depression, which contributes to the nonlinearity of $f(\texttt{dep}_i)$. In contrast, patients (\textbf{43}, \textbf{47}) show a healthy profile, with high emotional control, low depression, and no hormone deficiency nor use of growth hormone, prednisone, or hydrocortisone.

\section{Conclusions} \label{sec_conclu}
This study introduces the unit-Lindley additive partially linear model (UL-APLM), which extends the recently proposed unit-Lindley distribution to accommodate both linear and nonlinear covariate effects through penalized B-spline smoothing. The model provides a flexible, computationally efficient framework for analyzing bounded response variables, such as rates and proportions.

We develop the penalized likelihood estimation with the P-GAM algorithm, derive the effective degrees of freedom, and implement automatic smoothing parameter selection via Fellner–Schall updates. In addition, we derive and apply the local influence methodology to the UL-APLM, enabling robust diagnostic assessment under case-weight and response perturbations.

Comprehensive simulation studies demonstrate the model's consistency and stability, even in small samples, while a real application to psychological data highlights its interpretability and diagnostic power. Overall, this work contributes to the integration of modern additive regression and influence diagnostics within the unit-Lindley family, offering a new, tractable alternative to beta-based models for bounded data. All code used throughout the paper is provided in the supplementary materials for reproducibility.

\section*{Acknowledgements}
This work was partially supported by Conselho Nacional de Desenvolvimento Científico e Tecnológico (CNPq) - Brazil.

\section*{Supplementary material}
The supplementary material includes the following: \suppsection{sec_1deriv}{First derivatives}, \suppsection{sec_2deriv}{Second derivatives}, \suppsection{sec_Rgen}{Data generate R function}, \suppsection{sec_Rulaplm}{Fit model R function}, \suppsection{sec_Rsimu}{Simulation R commands}, \suppsection{sec_dataset}{Dataset}, \suppsection{sec_Rhypo}{Application R commands} and \suppsection{sec_Rdiag}{Diagnostic R tools}.

\bibliography{sn-biblio}



\section*{Supplementary material}

# Supplementary Material to: Inference and local influence diagnostics for unit-Lindley additive partially linear models

Hatice Tul Kubra Akdur[1], Danilo V. Silva[2], Gilberto A. Paula[2*]

[1]Department of Statistics, Faculty of Science, Gazi University, Ankara, Turkey.

[2]Department of Statistics, Institute of Mathematics, Statistics and Computer Science, Universidade de São Paulo, São Paulo, Brazil.

*Corresponding author(s). E-mail(s): giapaula@ime.usp.br;
Contributing authors: haticesenol@gazi.edu.tr; danilo.silva@ime.usp.br;

## 1 First derivatives

The mean response $\mu_i = g^{-1}(\eta_i)$ and with the linear predictor defined as

$$\eta_i = \mathbf{x}_i^{\mathrm{T}}\boldsymbol{\beta} + \sum_{j=1}^{r} \mathbf{N}_j(t_{ij})^{\mathrm{T}}\boldsymbol{\gamma}_j.$$

So, the derivative of the inverse link function is given by

$$g'(\mu_i) = \frac{dg(\mu_i)}{d\mu_i}, \quad \text{and} \quad \frac{d\mu_i}{d\eta_i} = \left(\frac{dg(\mu_i)}{d\mu_i}\right)^{-1},$$

the derivative of the log-likelihood function $\ell_i(\mu_i)$ with respect to $\mu_i$ is

$$\frac{d\ell_i}{d\mu_i} = \frac{-2}{1-\mu_i} - \frac{1}{\mu_i} + \frac{y_i}{\mu_i^2(1-y_i)},$$

and since $\mathbb{E}\left(d\ell_i/d\mu_i\right) = 0$ one has that

$$\mathbb{E}\left(\frac{Y_i}{1-Y_i}\right) = \frac{\mu_i(1+\mu_i)}{1-\mu_i}.$$

Using the chain rule, the penalized score function with respect to $\boldsymbol{\beta}$ becomes

$$\mathbf{U}^p_{\boldsymbol{\beta}} = \frac{\partial \ell_p}{\partial \boldsymbol{\beta}} = \sum_{i=1}^n \frac{d\ell_i}{d\mu_i} \cdot \frac{d\mu_i}{d\eta_i} \cdot \mathbf{x}_i = \sum_{i=1}^n \left(\frac{-2}{1-\mu_i} - \frac{1}{\mu_i} + \frac{y_i}{\mu_i^2(1-y_i)}\right)\left(\frac{d\mu_i}{d\eta_i}\right)\mathbf{x}_i,$$

similarly the penalized score function with respect to the spline coefficient vector $\boldsymbol{\gamma}_j$ is given by

$$\mathbf{U}^p_{\boldsymbol{\gamma}_j} = \frac{\partial \ell_p}{\partial \boldsymbol{\gamma}_j} = \sum_{i=1}^n \left(\frac{d\ell_i}{d\mu_i} \cdot \frac{d\mu_i}{d\eta_i} \cdot \mathbf{N}_j(t_{ij})\right) - \alpha_j \mathbf{M}_j \boldsymbol{\gamma}_j.$$

Let us define the weight corresponding to the first derivative of the regular log-likelihood function with respect to the linear predictor for the $i$th observation

$$\mathrm{W}_i^{(1)} = \left(\frac{d\ell_i}{d\mu_i} \cdot \frac{d\mu_i}{d\eta_i}\right).$$

Then, the full penalized score functions can be compactly written as

$$\mathbf{U}^p_{\boldsymbol{\beta}} = \mathbf{X}^{\mathrm{T}}\mathbf{W}^{(1)}\mathbf{1}_n, \quad \text{and} \quad \mathbf{U}^p_{\boldsymbol{\gamma}_j} = \mathbf{N}_j^{\mathrm{T}}\mathbf{W}^{(1)}\mathbf{1}_n - \alpha_j \mathbf{M}_j \boldsymbol{\gamma}_j,$$

where $\mathbf{X}$ is the $n \times p$ design matrix with rows $\mathbf{x}_i^{\mathrm{T}}$ , $\mathbf{N}_j$ is the $q_j \times n$ spline basis matrix with rows $\mathbf{N}_j(t_{ij})^{\mathrm{T}}$ for the covariate $t_j$, $\mathbf{W}^{(1)}$ is a diagonal matrix with elements $\mathrm{W}_i^{(1)}$, $i = 1, ..., n$, and $\mathbf{1}_n$ is a $n \times 1$ vector of ones. The MPLE is obtained by solving the system of equations $\mathbf{U}^p_{\boldsymbol{\beta}} = \mathbf{0}$ and $\mathbf{U}^p_{\boldsymbol{\gamma}_j} = \mathbf{0}$, for $j = 1, \ldots, r$. These equations are typically solved using a numerical optimization algorithm, such as Newton-Raphson or Fisher scoring, due to the nonlinear nature of the penalized log-likelihood function and the presence of spline-based components.

In particular, assuming the logit link function $g(\mu) = \mathrm{logit}(\mu)$, the inverse link function is $\mu_i = \{1 + \exp(-\eta_i)\}^{-1}$, and its derivative is $d\mu_i/d\eta_i = \mu_i(1-\mu_i)$. Thus, the full expression for the score weight becomes

$$\mathrm{W}_i^{(1)} = \left(\frac{-2}{1-\mu_i} - \frac{1}{\mu_i} + \frac{y_i}{\mu_i^2(1-y_i)}\right)\mu_i(1-\mu_i).$$

# 2 Second derivatives

According to the penalized log-likelihood, the second derivatives (Hessian components) with respect to $\boldsymbol{\beta}$ are given by

$$\frac{\partial^2 \ell_p}{\partial \boldsymbol{\beta}\, \partial \boldsymbol{\beta}^{\mathrm{T}}} = -\sum_{i=1}^n \mathrm{W}_i^{(2)}\, \mathbf{x}_i \mathbf{x}_i^{\mathrm{T}},$$

with respect to $\boldsymbol{\gamma}_j$ are

$$\frac{\partial^2 \ell_p}{\partial \boldsymbol{\gamma}_j \, \partial \boldsymbol{\gamma}_j^{\mathrm{T}}} = -\sum_{i=1}^{n} \mathrm{W}_i^{(2)} \, \mathbf{N}_j(t_{ij}) \mathbf{N}_j(t_{ij})^{\mathrm{T}} - \alpha_j \mathbf{M}_j,$$

the cross-partial derivatives of the penalized log-likelihood with respect to $\boldsymbol{\gamma}_j$ and $\boldsymbol{\gamma}_k$ are

$$\frac{\partial^2 \ell_p}{\partial \boldsymbol{\gamma}_j \, \partial \boldsymbol{\gamma}_k^{\mathrm{T}}} = -\sum_{i=1}^{n} \mathrm{W}_i^{(2)} \, \mathbf{N}_j(t_{ij}) \mathbf{N}_k(t_{ik})^{\mathrm{T}}, \quad \text{for } j \neq k,$$

the cross-derivative between $\boldsymbol{\beta}$ and $\boldsymbol{\gamma}_j$ are

$$\frac{\partial^2 \ell_p}{\partial \boldsymbol{\beta} \, \partial \boldsymbol{\gamma}_j^{\mathrm{T}}} = -\sum_{i=1}^{n} \mathrm{W}_i^{(2)} \, \mathbf{x}_i \mathbf{N}_j(t_{ij})^{\mathrm{T}},$$

where

$$\mathrm{W}_i^{(2)} = -\frac{d\mathrm{W}_i^{(1)}}{d\eta_i} = -\left\{ \left( \frac{d^2 \ell_i}{d\mu_i^2} \right) \left( \frac{1}{g'(\mu_i)} \right)^2 + \left( \frac{d\ell_i}{d\mu_i} \right) \left( -\frac{g''(\mu_i)}{(g'(\mu_i))^3} \right) \right\}.$$

The second derivative of the regular log-likelihood function $\ell_i(\mu_i)$ with respect to $\mu_i$ is given by

$$\frac{d^2 \ell_i}{d\mu_i^2} = -\frac{2}{(1-\mu_i)^2} + \frac{1}{\mu_i^2} - \frac{2y_i}{\mu_i^3 (1-y_i)},$$

and the closed-form expression for $\mathrm{W}_i^{(2)}$ becomes

$$\mathrm{W}_i^{(2)} = \left\{ \frac{2}{(1-\mu_i)^2} - \frac{1}{\mu_i^2} + \frac{2y_i}{\mu_i^3(1-y_i)} \right\} \left( \frac{1}{g'(\mu_i)} \right)^2 - \mathrm{W}_i^{(1)} \frac{g''(\mu_i)}{\{g'(\mu_i)\}^3}.$$

In particular, assuming the logit link function

$$g'(\mu) = \frac{1}{\mu(1-\mu)}, \quad \text{and} \quad g''(\mu) = \frac{2\mu - 1}{\mu^2(1-\mu)^2}.$$

Then, the penalized observed information matrix (negative Hessian) for $\boldsymbol{\theta}$ is defined as

$$\mathcal{I}_p^{(\boldsymbol{\theta},\boldsymbol{\theta})} = \begin{bmatrix} \mathcal{I}_p^{(\boldsymbol{\beta},\boldsymbol{\beta})} & \mathcal{I}_p^{(\boldsymbol{\beta},\boldsymbol{\gamma})} \\ \mathcal{I}_p^{(\boldsymbol{\gamma},\boldsymbol{\beta})} & \mathcal{I}_p^{(\boldsymbol{\gamma},\boldsymbol{\gamma})} \end{bmatrix},$$

with components with respect to $\boldsymbol{\beta}$ are

$$\mathcal{I}_p^{(\boldsymbol{\beta},\boldsymbol{\beta})} = \sum_{i=1}^{n} \mathrm{W}_i^{(2)} \, \mathbf{x}_i \mathbf{x}_i^{\mathrm{T}} = \mathbf{X}^{\mathrm{T}} \mathbf{W}^{(2)} \mathbf{X},$$

with respect to $\boldsymbol{\gamma}_j$

$$\mathcal{I}_p^{(\boldsymbol{\gamma}_j,\boldsymbol{\gamma}_j)} = \sum_{i=1}^{n} \mathrm{W}_i^{(2)} \, \mathbf{N}_j(t_{ij})\mathbf{N}_j(t_{ij})^{\mathrm{T}} + \alpha_j \mathbf{M}_j = \mathbf{N}_j^{\mathrm{T}} \mathbf{W}^{(2)} \mathbf{N}_j + \alpha_j \mathbf{M}_j,$$

cross-derivative between $\boldsymbol{\gamma}_j$ and $\boldsymbol{\gamma}_k$, $j \neq k$

$$\mathcal{I}_p^{(\boldsymbol{\gamma}_j,\boldsymbol{\gamma}_k)} = \sum_{i=1}^{n} \mathrm{W}_i^{(2)} \, \mathbf{N}_j(t_{ij})\mathbf{N}_k(t_{ik})^{\mathrm{T}} = \mathbf{N}_j^{\mathrm{T}} \mathbf{W}^{(2)} \mathbf{N}_k,$$

cross-derivative between $\boldsymbol{\beta}$ and $\boldsymbol{\gamma}_j$

$$\mathcal{I}_p^{(\boldsymbol{\beta},\boldsymbol{\gamma}_j)} = \sum_{i=1}^{n} \mathrm{W}_i^{(2)} \, \mathbf{x}_i \mathbf{N}_j(t_{ij})^{\mathrm{T}} = \mathbf{X}^{\mathrm{T}} \mathbf{W}^{(2)} \mathbf{N}_j,$$

where $\mathbf{W}^{(2)}$ is a diagonal matrix with entries $\mathrm{W}_i^{(2)}$.

# 3 Data generated R function

The R packages needed are `lamW` (Adler 2025), `Matrix` (Bates et al. 2025) and `splines` (Bates and Venables 1999). The `gen_data()` used in the simulation study is

```
> # packs
> library(lamW)
> library(Matrix)
> library(splines)
>
> # cdf
> pUL <- function(q, mu){
+   p <- numeric(length(q))
+   p <- 1-((1-mu*q)/(1-q))*exp(-q*(1-mu)/(mu*(1-q)))
+   p[q<=0] <- 0; p[q>=1] <- 1
+   p
+ }
>
> # quantile
> qUL <- function(p, mu){
+   n <- length(p)
+   q <- numeric(n)
+   for(i in 1:n){
+     f <- function(q) pUL(q, mu=mu[i]) - p[i]
+     q[i] <- uniroot(f, c(0,1), f.upper=1)$root
+   }
+   q
```

```
+ }
>
> # random generate
> rUL <- function(n, mu) qUL(runif(n), mu)
>
> # inverse link
> inv_link <- function(eta){
+  # return
+  thresh <- -log(.Machine$double.eps)
+  eta <- pmin(thresh, pmax(eta, -thresh))
+  1 / (1 + exp(-eta))
> }
>
> # generate dataset
> gen_data <- function(n, R){
+   # 1. covariates
+   set.seed(1234)
+   x1 <- rbinom(n, 1, 0.5)
+   x2 <- runif(n, -1, 1)
+   X <- model.matrix(~ 1 + x1 + x2)
+   dat <- list(X=X)
+
+   # 2. t1 and f1
+   t1 <- runif(n, -5, 0)
+   f1 <- cos(0.5 * pi * t1)
+   f1 <- f1 - mean(f1)
+   dat$t1 <- t1
+   dat$f1 <- f1
+
+   # 2. t2 and f2
+   t2 <- runif(n, 0, 3)
+   f2 <- (-t2^3 + exp(t2)/4)/5
+   f2 <- f2 - mean(f2)
+   dat$t2 <- t2
+   dat$f2 <- f2
+
+   # 3. mu
+   beta_true <- c(-0.5, 1, 0.5)
+   eta <- X %*% beta_true + f1 + f2
+   mu <- inv_link(eta)
+   dat$mu <- as.numeric(mu)
+   dat
+
+   # 4. b-spline B1
+   k1 <- 8
```

```
+   knots <- seq(min(dat$t1), max(dat$t1), length.out=k1+2)
+   knots <- knots[-c(1, k1 + 2)]
+   knots <- c(rep(min(dat$t1), 4), knots, rep(max(dat$t1), 4))
+   B1 <- splineDesign(dat$t1, knots=knots, ord=4, outer.ok=T)
+   B1 <- scale(B1, scale=F)
+   B1 <- B1[,-ncol(B1)]
+   dat$B1 <- B1
+
+   # 4. b-spline B2
+   k2 <- 8
+   knots <- seq(min(dat$t2), max(dat$t2), length.out=k2+2)
+   knots <- knots[-c(1, k2 + 2)]
+   knots <- c(rep(min(dat$t2), 4), knots, rep(max(dat$t2), 4))
+   B2 <- splineDesign(dat$t2, knots=knots, ord=4, outer.ok=T)
+   B2 <- scale(B2, scale=F)
+   B2 <- B2[,-ncol(B2)]
+   dat$B2 <- B2
+
+   # 5. penalty matrix
+   M1 <- get_penalty(ncol(B1))
+   M2 <- get_penalty(ncol(B2))
+   dat$M1 <- M1
+   dat$M2 <- M2
+
+   # 6. random y
+   for(r in 1:R){
+     # rUL
+     set.seed(r)
+     yr <- rUL(n, mu)
+     dat[[paste0("y", r)]] <- yr
+   }
+   dat
+ }
```

# 4 Fit model R function

The additional R packages needed are `MASS` (Ripley et al. 2025) and `glmnet` (Friedman et al. 2025). The `fit_ulaplm()` used in the application analysis is

```
> # packs
> library(MASS)
> library(glmnet)
>
> # dmu/deta
> dmu_deta <- function(mu) mu * (1-mu)
```

```
>
> # d2mu/deta2
> d2mu_deta2 <- function(mu) dmu_deta(mu) * (1-2*mu)
>
> # score
> dell_dmu <- function(mu, y) -2/(1-mu) - 1/mu + y/(mu^2*(1-y))
>
> # score second
> d2ell_dmu2 <- function(mu, y) -2/((1-mu)^2) + 1/(mu^2) -2*y/(mu^3*(1-y))
>
> # first weight
> get_W1 <- function(mu, y) as.vector(dell_dmu(mu, y) * dmu_deta(mu))
>
> # second weight
> get_W2 <- function(mu, y){
+   term1 <- d2ell_dmu2(mu, y) * (dmu_deta(mu)^2)
+   term2 <- dell_dmu(mu, y) * d2mu_deta2(mu)
+   as.vector(-(term1 + term2))
+ }
>
> # third weight (response perturbation)
> get_W3 <- function(mu, y) as.vector((1-mu) / (mu*(1-y)^2))
>
> # penalty matrix (reparam)
> get_penalty <- function(k) {
+   D <- diff(diag(k+1), differences = 2)
+   D <- D[,-(k+1)]
+   t(D) %*% D
+ }
>
> # info_matrix
> get_info <- function(N, mu, y, M){
+   # information
+   I_full <- t(N) %*% (get_W2(mu, y) * N) + M
+   I_full
+ }
>
> # update theta
> update_theta <- function(theta_old, N, y, M){
+   # mu
+   mu <- inv_link(N%*%theta_old)
+
+   # score (gradient)
+   U_full <- t(N) %*% get_W1(mu, y) - M %*% theta_old
+
+   # observed Information
```

```
+   I_mat <- get_info(N, mu, y, M)
+
+   # inverse
+   L <- chol(I_mat)
+   b <- forwardsolve(t(L), U_full)
+   theta_new <- backsolve(L, b) + theta_old
+   theta_new
+ }
>
> # update alpha
> update_alpha <- function(theta_new, N, y, M, M_list, alpha_old, gama_pos){
+   # mu
+   mu <- inv_link(N%*%theta_new)
+
+   # observed Information
+   I_mat <- get_info(N, mu, y, M)
+
+   # alpha new
+   r <- length(M_list)
+   alpha_new <- alpha_old
+   for(j in 1:r){
+     # prep
+     ini_j <- gama_pos$ini[j]
+     fin_j <- gama_pos$fin[j]
+     Mbig_j <- matrix(0, ncol(N), ncol(N))
+     Mbig_j[ini_j:fin_j, ini_j:fin_j] <- M_list[[j]]
+
+     # prep
+     tr1 <- sum(diag(ginv(as.matrix(M))%*%Mbig_j))
+     tr2 <- sum(diag(solve(I_mat, Mbig_j)))
+
+     # update
+     gama_j <- theta_new[ini_j:fin_j]
+     alpha_new[j] <- alpha_new[j] / t(gama_j) %*% M_list[[j]] %*% gama_j
+     alpha_new[j] <- (tr1 - tr2)*alpha_new[j]
+   }
+   alpha_new
+ }
>
> # fit
> fit_ulaplm <- function(X, B_list, y, M_list){
+   # initial alpha
+   p <- ncol(X)
+   r <- length(B_list)
+   alpha_old <- runif(r, 1, 1000)
+
```

```
+   # compute N
+   N <- X
+   theta_ini <- 1
+   theta_fin <- p
+   for(j in 1:r){
+     N <- cbind(N, B_list[[j]])
+     theta_ini <- c(theta_ini, tail(theta_fin,1)+1)
+     theta_fin <- c(theta_fin, tail(theta_fin,1)+ncol(B_list[[j]]))
+   }
+   gama_pos <- list(ini=theta_ini[-1], fin=theta_fin[-1])
+
+   # initial theta
+   theta_old <- coef(glmnet(N[,-1], inv_link(y), alpha=0, lambda=10))
+   theta_old <- as.vector(theta_old)
+
+   # fit
+   sair <- 0; i <- 0
+   while(sair==0){
+     # compute M_alpha
+     M <- diag(p)*0
+     for(j in 1:r) M <- bdiag(M, alpha_old[j]*M_list[[j]])
+
+     # update theta
+     theta_new <- update_theta(theta_old, N, y, M)
+     dif <- mean(abs((theta_new-theta_old)/theta_old))
+     dif <- round(dif, 4)
+
+     # update alpha
+     alpha_new <- update_alpha(theta_new, N, y, M,
+                               M_list, alpha_old, gama_pos)
+
+     # check
+     if(dif < 1e-3) sair <- 1
+     if(sair == 0) alpha_old <- alpha_new
+     theta_old <- theta_new
+     i <- i+1
+   }
+
+   # final
+   cat("\r", "Iteration:", i, "-", "The metric is:", dif, "\n")
+   fit <- list()
+   fit$theta <- theta_old
+   fit$mu <- inv_link(N%*%fit$theta)
+   fit$y <- y
+   fit$N <- N
```

```
+   fit$alpha <- alpha_old
+
+   # edf + vcov
+   vcov <- qr.solve(qr(get_info(N, fit$mu, y, M)))
+   fit$vcov <- vcov
+   H <- vcov %*% t(N) %*% (get_W2(fit$mu, y) * N)
+   fit$edf <- diag(H)
+
+   # sep coefs
+   fit$beta <- fit$theta[1:p]
+   r <- length(M_list)
+   for(j in 1:r){
+     # fj
+     ini_j <- gama_pos$ini[j]
+     fin_j <- gama_pos$fin[j]
+     fj <- B_list[[j]] %*% fit$theta[ini_j:fin_j]
+     fit[[paste0("f", j)]] <- fj
+
+     # fj_upp + fj_low
+     fj_vcov <- vcov[ini_j:fin_j, ini_j:fin_j]
+     fj_vcov <- B_list[[j]] %*% fj_vcov %*% t(B_list[[j]])
+     fit[[paste0("f", j, "_upp")]] <- fj + qnorm(0.975)*sqrt((diag(fj_vcov)))
+     fit[[paste0("f", j, "_low")]] <- fj - qnorm(0.975)*sqrt((diag(fj_vcov)))
+   }
+   fit
+ }
```

# 5 Simulation R commands

The simulation study is performed with

```
> # load fit_ulaplm
> source("R/ulaplm.R")
>
> # simu
> R <- 5
> n1 <- 50
> n2 <- 200
> n3 <- 800
> dat_n1 <- gen_data(n1, R)
> dat_n2 <- gen_data(n2, R)
> dat_n3 <- gen_data(n3, R)
>
> # fit for n1
> fit_n1 <- list()
```

```
> for(r in 1:R){
+   # prep
+   X <- dat_n1$X
+   B_list <- list(dat_n1$B1, dat_n1$B2)
+   yr <- dat_n1[[paste0("y", r)]]
+   M_list <- list(dat_n1$M1, dat_n1$M2)
+
+   # model
+   fitr <- fit_ulaplm(X, B_list, yr, M_list)
+   fit_n1[[paste0("fit", r)]] <- fitr
+ }
>
> # fit for n2
> fit_n2 <- list()
> for(r in 1:R){
+   # prep
+   X <- dat_n2$X
+   B_list <- list(dat_n2$B1, dat_n2$B2)
+   yr <- dat_n2[[paste0("y", r)]]
+   M_list <- list(dat_n2$M1, dat_n2$M2)
+
+   # model
+   fitr <- fit_ulaplm(X, B_list, yr, M_list)
+   fit_n2[[paste0("fit", r)]] <- fitr
+ }
>
> # fit for n3
> fit_n3 <- list()
> for(r in 1:R){
+   # prep
+   X <- dat_n3$X
+   B_list <- list(dat_n3$B1, dat_n3$B2)
+   yr <- dat_n3[[paste0("y", r)]]
+   M_list <- list(dat_n3$M1, dat_n3$M2)
+
+   # model
+   fitr <- fit_ulaplm(X, B_list, yr, M_list)
+   fit_n3[[paste0("fit", r)]] <- fitr
+ }
```

# 6 Dataset

The data set analyzed from the Center of Applied Statistics of the University of São Paulo (Paula et al. 2023) on the assessment of the psychological profile of patients with hypopituitarism follows

```
> # dataset
> dat
   id resp dep adreno drug1 drug2
1   1 0.80   1     no    no  none
2   2 0.01  27    yes    no  none
3   3 0.30  34    yes    no  none
4   4 0.70  22    yes   yes predi
5   5 0.10  90     no   yes hidro
6   6 0.10  59     no   yes hidro
7   7 0.01  55     no    no  none
8   8 0.10  75    yes   yes predi
9   9 0.20  25     no   yes hidro
10 10 0.50  12    yes   yes predi
11 11 0.70  82    yes   yes hidro
12 12 0.10  63    yes   yes hidro
13 13 0.40  63     no    no  none
14 14 0.50  10     no   yes predi
15 15 0.30  39    yes   yes predi
16 16 0.50   5     no   yes hidro
17 17 0.70  74     no   yes predi
18 18 0.20  80     no    no  none
19 19 0.40  36     no    no  none
20 20 0.20  72     no   yes hidro
21 21 0.01  82     no   yes predi
22 22 0.10  28    yes    no  none
23 23 0.60  16     no    no  none
24 24 0.50  55    yes    no  none
25 25 0.60  97     no    no  none
26 26 0.60   2     no    no  none
27 27 0.50  38     no    no  none
28 28 0.40   2     no    no  none
29 29 0.10  81    yes   yes hidro
30 30 0.50  10    yes   yes  none
31 31 0.70  13     no   yes predi
32 32 0.10  14     no   yes hidro
33 33 0.80   5    yes   yes  none
34 34 0.20  29    yes   yes predi
35 35 0.50   5     no    no hidro
36 36 0.80  12    yes   yes predi
37 37 0.70  24     no   yes hidro
38 38 0.70  63     no   yes predi
39 39 0.01  34    yes    no  none
40 40 0.30  24     no   yes hidro
41 41 0.10  69    yes   yes hidro
42 42 0.01  51    yes    no  none
43 43 0.90   2     no    no  none
```

```
44 44 0.01  85     no    no  none
45 45 0.70  83    yes   yes  none
46 46 0.10  17    yes   yes hidro
47 47 0.90  16     no    no  none
48 48 0.40  27     no   yes hidro
49 49 0.80   3     no   yes hidro
50 50 0.50   2     no   yes predi
51 51 0.50  76     no    no  none
52 52 0.60   1     no   yes predi
53 53 0.01  50    yes   yes predi
54 54 0.50  14    yes    no  none
```

# 7 Application R commands

The application analysis is performed with

```
> # load fit_ulaplm
> source("R/ulaplm.R")
>
> # prep data
> dat1 <- read.table("data/hypo.txt", header=T)
> X <- model.matrix(~1+adreno+drug1+drug2, data=dat1)
> y <- dat1$resp
>
> # b-spline
> k1 <- 8
> knots <- seq(min(dat1$dep), max(dat1$dep), length.out=k1+2)
> knots <- knots[-c(1, k1 + 2)]
> knots <- c(rep(min(dat1$dep), 4), knots, rep(max(dat1$dep), 4))
> B1 <- splineDesign(dat1$dep, knots=knots, ord=4, outer.ok=T)
> B1 <- scale(B1, scale=F)
> B1 <- B1[,-ncol(B1)]
> B_list <- list(B1)
>
> # penalty matrix
> M1 <- get_penalty(ncol(B1))
> M_list <- list(M1)
>
> # model
> b1 <- fit_ulaplm(X, B_list, y, M_list)
 Iteration: 14 - The metric is: 9e-04
>
> # summary
> b1$beta
[1] -0.09858488 -0.44272374  0.67273834 -0.52646846 -1.02343286
```

```
>
> sqrt(diag(b1$vcov)[1:5])
[1] 0.1821001 0.2471068 0.4314358 0.4452977 0.4151965
>
> b1$alpha
[1] 21.06709
>
> sum(b1$edf)
[1] 7.796897
>
> # test
> w <- abs(b1$beta/sqrt(diag(b1$vcov)[1:5]))
> 1 - pnorm(w)
[1] 0.294123784 0.036596203 0.059462562 0.118546453 0.006851882
>
> # test
> I_mat <- qr.solve(qr(b1$vcov))
> w <- t(b1$theta[-(1:5)]) %*% I_mat[-(1:5),-(1:5)] %*% b1$theta[-(1:5)]
> 1 - pchisq(as.numeric(w), sum(b1$edf[-(1:5)]))
[1] 9.914006e-05
>
> # residual
> res_ulaplm(b1)
>
> # local influ
> Bi_ulaplm(b1, c(2,2))
> dmax_ulaplm(b1, c(1,3))
>
> # model (drop 11)
> dat2 <- read.table("data/hypo.txt", header=T)
> dat2 <- dat2[-c(11),]
> X <- model.matrix(~1+adreno+drug1+drug2, data=dat2)
> y <- dat2$resp
> k1 <- 8
> knots <- seq(min(dat2$dep), max(dat2$dep), length.out=k1+2)
> knots <- knots[-c(1, k1 + 2)]
> knots <- c(rep(min(dat2$dep), 4), knots, rep(max(dat2$dep), 4))
> B1 <- splineDesign(dat2$dep, knots=knots, ord=4, outer.ok=T)
> B1 <- scale(B1, scale=F)
> B1 <- B1[,-ncol(B1)]
> B_list <- list(B1)
> M1 <- get_penalty(ncol(B1))
> M_list <- list(M1)
> b2 <- fit_ulaplm(X, B_list, y, M_list)
 Iteration: 18 - The metric is: 6e-04
>
```

```
> # model (drop 43 and 47)
> dat3 <- read.table("data/hypo.txt", header=T)
> dat3 <- dat3[-c(43, 47),]
> X <- model.matrix(~1+adreno+drug1+drug2, data=dat3)
> y <- dat3$resp
> k1 <- 8
> knots <- seq(min(dat3$dep), max(dat3$dep), length.out=k1+2)
> knots <- knots[-c(1, k1 + 2)]
> knots <- c(rep(min(dat3$dep), 4), knots, rep(max(dat3$dep), 4))
> B1 <- splineDesign(dat3$dep, knots=knots, ord=4, outer.ok=T)
> B1 <- scale(B1, scale=F)
> B1 <- B1[,-ncol(B1)]
> B_list <- list(B1)
> M1 <- get_penalty(ncol(B1))
> M_list <- list(M1)
> b3 <- fit_ulaplm(X, B_list, y, M_list)
 Iteration: 7 - The metric is: 6e-04
```

# 8 Diagnostic R tools

The `res_ulaplm()`, `Bi_ulaplm()` and `dmax_ulaplm()` used in the application analysis are

```
> # residual plot
> res_ulaplm <- function(fit){
+   # quantile residual
+   rq <- qnorm(pUL(fit$y, fit$mu))
+
+   # rq vs mu
+   par(mar=c(4.5,5,1,1), mfrow=c(1,3), pch=19, cex.lab=1.8, cex.axis=1.5)
+   plot(rq~fit$mu, xlab="fitted value", ylab="quantile residual")
+   lines(smooth.spline(y=rq, fit$mu, df=4), lwd=2)
+
+   # rq qqplot
+   qqnorm(rq, xlab="theoretical quantile", ylab="quantile residual", main="")
+   abline(b=1, a=0, lwd=2, lty=2)
+
+   # rq wormplot
+   rq_theor <- qnorm(ppoints(length(rq)))
+   rq_cent <- sort(rq) - rq_theor
+   p <- pnorm(rq_theor)
+   se <- (1/dnorm(rq_theor)) * sqrt(p*(1-p)/length(rq))
+   conf_upp <- qnorm(0.975)*se
+   conf_low <- qnorm(0.025)*se
+   plot(rq_cent~rq_theor, xlab="theoretical quantile",
```

```
+        ylab="centered quantile residual", ylim=c(-1,1))
+   lines(smooth.spline(y=rq_cent, rq_theor, df=4), lwd=2)
+   lines(conf_upp~rq_theor, lwd=2, lty=2)
+   lines(conf_low~rq_theor, lwd=2, lty=2)
+ }
>
> # local influence
> Bi_ulaplm <- function(fit, m=rep(0,2)){
+   # cw perturbation
+   Delta_cw <- t(fit$N * get_W1(fit$mu, fit$y))
+   B_cw <- t(Delta_cw) %*% fit$vcov %*% Delta_cw
+   Bi_cw <- diag(B_cw)/sqrt(sum(diag(B_cw%*%B_cw)))
+
+   # rp perturbation
+   Delta_rp <- t(fit$N * get_W3(fit$mu, fit$y))
+   B_rp <- t(Delta_rp) %*% fit$vcov %*% Delta_rp
+   Bi_rp <- diag(B_rp)/sqrt(sum(diag(B_rp%*%B_rp)))
+   measu <- list(Bi_cw=Bi_cw, Bi_rp=Bi_rp)
+
+   # prep plot
+   labs <- c(expression("B"["i"] ~ ": case-weight pertubation"),
+             expression("B"["i"] ~ ": response pertubation"))
+   par(mar=c(4.5,5,1,1), mfrow=c(1,2), pch=19, cex.lab=1.3, cex.axis=1)
+
+   # plot
+   n <- nrow(fit$N)
+   for(i in 1:2){
+     cut <- mean(measu[[i]]) + 2*sd(measu[[i]])
+     plot(measu[[i]], x=1:n, ylab=labs[i], xlab="index")
+     abline(h=cut, lty=2, lwd=2)
+     if(m[i]>0) identify(x=1:n, y=measu[[i]], n=m[i], cex=1)
+   }
+ }
>
> # local influence
> dmax_ulaplm <- function(fit, m=rep(0,2)){
+   # cw perturbation
+   Delta_cw <- t(fit$N * get_W1(fit$mu, fit$y))
+   B_cw <- t(Delta_cw) %*% fit$vcov %*% Delta_cw
+   eig_cw <- eigen(B_cw, symmetric=T)
+   dmax_cw <- abs(eig_cw$vectors[1,])
+
+   # rp perturbation
+   Delta_rp <- t(fit$N * get_W3(fit$mu, fit$y))
+   B_rp <- t(Delta_rp) %*% fit$vcov %*% Delta_rp
```

```
+   eig_rp <- eigen(B_rp, symmetric=T)
+   dmax_rp <- abs(eig_rp$vectors[1,])
+   measu <- list(dmax_cw=dmax_cw, dmax_rp=dmax_rp)
+
+   # prep plot
+   labs <- c(expression("dmax"["i"] ~ ": case-weight pertubation"),
+             expression("dmax"["i"] ~ ": response pertubation"))
+   par(mar=c(4.5,5,1,1), mfrow=c(1,2), pch=19, cex.lab=1.3, cex.axis=1)
+
+   # plot
+   n <- nrow(fit$N)
+   for(i in 1:2){
+     cut <- mean(measu[[i]]) + 2*sd(measu[[i]])
+     plot(measu[[i]], x=1:n, ylab=labs[i], xlab="index")
+     abline(h=cut, lty=2, lwd=2)
+     if(m[i]>0) identify(x=1:n, y=measu[[i]], n=m[i], cex=1)
+   }
+ }
```



\end{document}